\documentclass[aps,prb,twocolumn,superscriptaddress]{revtex4-1}
\usepackage{blindtext}
\usepackage{amsmath,amssymb,amsfonts}
\usepackage{bbm}
\usepackage{bm}
\usepackage{color}
\usepackage[dvipsnames]{xcolor}
\usepackage{stackengine}
\usepackage{epstopdf}
\usepackage[normalem]{ulem} 
\usepackage{soul} 

\usepackage{graphicx}
\usepackage{indentfirst}
\usepackage{psfrag}
\usepackage{epsfig}

\newcommand{\bra}[1]{\ensuremath{\left\langle#1\right|}}
\newcommand{\ket}[1]{\ensuremath{\left|#1\right\rangle}}

\DeclareMathOperator{\Tr}{Tr}

\renewcommand{\Re}{\mathrm{Re}}
\renewcommand{\Im}{\mathrm{Im}}
\newcommand{\screen}{\ensuremath{\phi}}

\def\tmp{%
\begin{bmatrix}
	\langle \mathbf{L}_{om0}^{(1)}(k_{1l}), \mathbf{L}_{om0}^{(1)}(k_{2l})\rangle & 
	\langle \mathbf{L}_{om0}^{(1)}(k_{1l}), \mathbf{N}_{l,om0}^{(1)}(k_{2s})\rangle & \dots & 
	\langle \mathbf{L}_{om0}^{(1)}(k_{1l}), \mathbf{M}_{l,emN}^{(3)}(k_{1s})\rangle \\
	\langle \mathbf{N}_{l,om0}^{(1)}(k_{1s}), \mathbf{L}_{om0}^{(1)}(k_{2l})\rangle & 
	\langle \mathbf{N}_{l,om0}^{(1)}(k_{1s}), \mathbf{N}_{l,om0}^{(1)}(k_{2s})\rangle & \dots & 
	\langle \mathbf{N}_{l,om0}^{(1)}(k_{1s}), \mathbf{M}_{l,emN}^{(3)}(k_{1s})\rangle \\
	\hdotsfor{4} \\
	\langle \mathbf{M}_{l,emN}^{(1)}(k_{1s}), \mathbf{L}_{omN}^{(1)}(k_{2l})\rangle & 
	\langle \mathbf{M}_{l,emN}^{(1)}(k_{1s}), \mathbf{N}_{l,omN}^{(1)}(k_{2s})\rangle & \dots & 
	\langle \mathbf{M}_{l,emN}^{(1)}(k_{1s}), \mathbf{M}_{l,emN}^{(3)}(k_{1s})\rangle
\end{bmatrix}
}%

\def\tmpS{%
	\begin{bmatrix}
		\langle \mathbf{L}_{om0}^{(1)}(k_{1l}), \mathcal{S}_r[\mathbf{L}_{om0}^{(1)}(k_{2l})]\rangle & 
		\langle \mathbf{L}_{om0}^{(1)}(k_{1l}), \mathcal{S}_r[\mathbf{N}_{l,om0}^{(1)}(k_{2s})]\rangle & \dots & 
		\langle \mathbf{L}_{om0}^{(1)}(k_{1l}), \mathcal{S}_r[\mathbf{M}_{l,emN}^{(3)}(k_{1s})]\rangle \\
		\langle \mathbf{N}_{l,om0}^{(1)}(k_{1s}), \mathcal{S}_r[\mathbf{L}_{om0}^{(1)}(k_{2l})]\rangle & 
		\langle \mathbf{N}_{l,om0}^{(1)}(k_{1s}), \mathcal{S}_r[\mathbf{N}_{l,om0}^{(1)}(k_{2s})]\rangle & \dots & 
		\langle \mathbf{N}_{l,om0}^{(1)}(k_{1s}), \mathbf{M}_{l,emN}^{(3)}(k_{1s})\rangle \\
		\hdotsfor{4} \\
		\langle \mathbf{M}_{l,emN}^{(1)}(k_{1s}), \mathcal{S}_r[\mathbf{L}_{omN}^{(1)}(k_{2l})]\rangle & 
		\langle \mathbf{M}_{l,emN}^{(1)}(k_{1s}), \mathcal{S}_r[\mathbf{N}_{l,omN}^{(1)}(k_{2s})]\rangle & \dots & 
		\langle \mathbf{M}_{l,emN}^{(1)}(k_{1s}), \mathcal{S}_r[\mathbf{M}_{l,emN}^{(3)}(k_{1s})]\rangle
	\end{bmatrix}
}%

\def\tmpboth{%
	\begin{bmatrix}
		\langle \mathbf{L}_{om0}^{(1)}(k_{1l}), \mathbf{L}_{om0}^{(1)}(k_{2l})\rangle & 
		\langle \mathbf{L}_{om0}^{(1)}(k_{1l}), \mathbf{N}_{l,om0}^{(1)}(k_{2s})\rangle & \dots & 
		\langle \mathbf{L}_{om0}^{(1)}(k_{1l}), \mathbf{M}_{l,emN}^{(3)}(k_{1s})\rangle \\
		\langle \mathbf{N}_{l,om0}^{(1)}(k_{1s}), \mathbf{L}_{om0}^{(1)}(k_{2l})\rangle & 
		\langle \mathbf{N}_{l,om0}^{(1)}(k_{1s}), \mathbf{N}_{l,om0}^{(1)}(k_{2s})\rangle & \dots & 
		\langle \mathbf{N}_{l,om0}^{(1)}(k_{1s}), \mathbf{M}_{l,emN}^{(3)}(k_{1s})\rangle \\
		\hdotsfor{4} \\
		\langle \mathbf{M}_{l,emN}^{(1)}(k_{1s}), \mathbf{L}_{omN}^{(1)}(k_{2l})\rangle & 
		\langle \mathbf{M}_{l,emN}^{(1)}(k_{1s}), \mathbf{N}_{l,omN}^{(1)}(k_{2s})\rangle & \dots & 
		\langle \mathbf{M}_{l,emN}^{(1)}(k_{1s}), \mathbf{M}_{l,emN}^{(3)}(k_{1s})\rangle \\
		\langle \mathbf{L}_{om0}^{(1)}(k_{1l}), \mathcal{S}_r[\mathbf{L}_{om0}^{(1)}(k_{2l})]\rangle & 
		\langle \mathbf{L}_{om0}^{(1)}(k_{1l}), \mathcal{S}_r[\mathbf{N}_{l,om0}^{(1)}(k_{2s})]\rangle & \dots & 
		\langle \mathbf{L}_{om0}^{(1)}(k_{1l}), \mathcal{S}_r[\mathbf{M}_{l,emN}^{(3)}(k_{1s})]\rangle \\
		\langle \mathbf{N}_{l,om0}^{(1)}(k_{1s}), \mathcal{S}_r[\mathbf{L}_{om0}^{(1)}(k_{2l})]\rangle & 
		\langle \mathbf{N}_{l,om0}^{(1)}(k_{1s}), \mathcal{S}_r[\mathbf{N}_{l,om0}^{(1)}(k_{2s})]\rangle & \dots & 
		\langle \mathbf{N}_{l,om0}^{(1)}(k_{1s}), \mathbf{M}_{l,emN}^{(3)}(k_{1s})\rangle \\
		\hdotsfor{4} \\
		\langle \mathbf{M}_{l,emN}^{(1)}(k_{1s}), \mathcal{S}_r[\mathbf{L}_{omN}^{(1)}(k_{2l})]\rangle & 
		\langle \mathbf{M}_{l,emN}^{(1)}(k_{1s}), \mathcal{S}_r[\mathbf{N}_{l,omN}^{(1)}(k_{2s})]\rangle & \dots & 
		\langle \mathbf{M}_{l,emN}^{(1)}(k_{1s}), \mathcal{S}_r[\mathbf{M}_{l,emN}^{(3)}(k_{1s})]\rangle
	\end{bmatrix}
}%

\def\tmpp{%
	\begin{bmatrix}
		\langle \mathbf{L}_{om0}^{(1)}(k_{1l}), \mathbf{L}_{om0}^{(1)}(k_{1l})\rangle & 
		\langle \mathbf{L}_{om0}^{(1)}(k_{1l}), \mathbf{N}_{l,om0}^{(1)}(k_{1s})\rangle & \dots & 
		\langle \mathbf{L}_{om0}^{(1)}(k_{1l}), \mathbf{M}_{l,emN}^{(1)}(k_{1s})\rangle \\
		\langle \mathbf{N}_{l,om0}^{(1)}(k_{1s}), \mathbf{L}_{om0}^{(1)}(k_{1l})\rangle & 
		\langle \mathbf{N}_{l,om0}^{(1)}(k_{1s}), \mathbf{N}_{l,om0}^{(1)}(k_{1s})\rangle & \dots & 
		\langle \mathbf{N}_{l,om0}^{(1)}(k_{1s}), \mathbf{M}_{l,emN}^{(1)}(k_{1s})\rangle \\
		\hdotsfor{4} \\
		\langle \mathbf{M}_{l,emN}^{(1)}(k_{1s}), \mathbf{L}_{omN}^{(1)}(k_{1l})\rangle & 
		\langle \mathbf{M}_{l,emN}^{(1)}(k_{1s}), \mathbf{N}_{l,omN}^{(1)}(k_{1s})\rangle & \dots & 
		\langle \mathbf{M}_{l,emN}^{(1)}(k_{1s}), \mathbf{M}_{l,emN}^{(1)}(k_{1s})\rangle
	\end{bmatrix}
}%

\def\tmppS{%
	\begin{bmatrix}
		\langle \mathbf{L}_{om0}^{(1)}(k_{1l}), \mathcal{S}_r[\mathbf{L}_{om0}^{(1)}(k_{1l})]\rangle & 
		\langle \mathbf{L}_{om0}^{(1)}(k_{1l}), \mathcal{S}_r[\mathbf{N}_{l,om0}^{(1)}(k_{1s})]\rangle & \dots & 
		\langle \mathbf{L}_{om0}^{(1)}(k_{1l}), \mathcal{S}_r[\mathbf{M}_{l,emN}^{(1)}(k_{1s})]\rangle \\
		\langle \mathbf{N}_{l,om0}^{(1)}(k_{1s}), \mathcal{S}_r[\mathbf{L}_{om0}^{(1)}(k_{1l})]\rangle & 
		\langle \mathbf{N}_{l,om0}^{(1)}(k_{1s}), \mathcal{S}_r[\mathbf{N}_{l,om0}^{(1)}(k_{1s})]\rangle & \dots & 
		\langle \mathbf{N}_{l,om0}^{(1)}(k_{1s}), \mathcal{S}_r[\mathbf{M}_{l,emN}^{(1)}(k_{1s})]\rangle \\
		\hdotsfor{4} \\
		\langle \mathbf{M}_{l,emN}^{(1)}(k_{1s}), \mathcal{S}_r[\mathbf{L}_{omN}^{(1)}(k_{1l})]\rangle & 
		\langle \mathbf{M}_{l,emN}^{(1)}(k_{1s}), \mathcal{S}_r[\mathbf{N}_{l,omN}^{(1)}(k_{1s})]\rangle & \dots & 
		\langle \mathbf{M}_{l,emN}^{(1)}(k_{1s}), \mathcal{S}_r[\mathbf{M}_{l,emN}^{(1)}(k_{1s})]\rangle
	\end{bmatrix}
}%

\begin{document}
	\title{
		Elastic Purcell Effect
	}
	\author{Miko\l{}aj K. Schmidt}
	\email{mikolaj.schmidt@mq.edu.au}
	\affiliation{Centre for Ultrahigh bandwidth Devices for Optical Systems (CUDOS), Australia}
	\affiliation{Macquarie University Research Centre in Quantum Science and Technology (QSciTech), MQ Photonics Research Centre, Department of Physics and Astronomy, Macquarie University, NSW 2109, Australia.}
	\author{L.G. Helt}
	\affiliation{Department of Physics, Engineering Physics \& Astronomy, Queen's University, Kingston, ON, K7L 3N6, Canada.}
	\author{Christopher G. Poulton}
	\affiliation{Centre for Ultrahigh bandwidth Devices for Optical Systems (CUDOS), Australia}
	\affiliation{School of Mathematical and Physical Sciences, University of Technology Sydney, NSW 2007, Australia.}
	\author{M.J. Steel}
	\affiliation{Centre for Ultrahigh bandwidth Devices for Optical Systems (CUDOS), Australia}
	\affiliation{Macquarie University Research Centre in Quantum Science and Technology (QSciTech), MQ Photonics Research Centre, Department of Physics and Astronomy, Macquarie University, NSW 2109, Australia.}
	
	\begin{abstract}
		In this work, we introduce an elastic analog of the Purcell effect and show theoretically that spherical nanoparticles can serve as tunable and robust antennas for modifying the emission from localized elastic sources. This effect can be qualitatively described by introducing elastic counterparts of the familiar electromagnetic parameters: local density of elastic states, elastic Purcell factor, and effective volume of elastic modes. To illustrate our framework, we consider the example of a submicron gold sphere as a generic elastic GHz antenna and find that shear and mixed modes of low orders in such systems offer considerable elastic Purcell factors. This formalism opens pathways towards extended control over dissipation of vibrations in various optomechanical systems and contributes to closing the gap between classical and quantum-mechanical treatments of phonons localized in elastic nanoresonators.
	\end{abstract}		
	\maketitle

In 1946,\cite{Purcell1946} E.M. Purcell proposed that the rate of nuclear magnetic moment transitions can be increased by coupling the system to an electric circuit. This enhancement, the \textit{Purcell effect}, depends on the fundamental properties of the circuit or, more generally, the mode to which the system is coupled: its resonant frequency, volume ($V$) and quality factor ($Q$). Ever since, considerable effort has been devoted to designing tunable resonators with ever smaller mode volumes and larger $Q$s. Recently, a similar modification of the decay rate has been demonstrated in the acoustic domain by measuring changes in the damping of oscillations of a Chinese gong placed near a hard wall.\cite{langguth2016drexhage} While that system does not exhibit any resonant behavior, or a clear cavity-like structure, it suggests that the original idea from Purcell can be extended to the physics of mechanical vibrations. In enabling such a direct analog, we provide a robust framework for calculating, and ultimately engineering, the output of an elastic source. The ability to control the rate of mechanical dissipation can be extended to spatial and spectral degrees of freedom by engineering the local density of elastic states (LDES) with resonant elastic nanosystems, and leads to the possibility of manipulating phonon-mediated nonlinear optical processes.\cite{eichenfield2009optomechanical,sipe2016hamiltonian,merklein2017chip,behunin2017engineering}

In this Letter, we investigate the elastic Purcell factor by analyzing the response of elastic antennas formed by spherical nanoparticles, and discuss how they can enhance or quench the emission from nearby sources of elastic waves.
To {go beyond purely acoustic phenomena,\cite{faran1951sound,langguth2016drexhage} we} adopt the theory of elastic wave scattering\cite{ying1956scattering,einspruch1960scattering,pao1973diffraction,achenbach2012wave,graff2012wave} which accounts for the presence of both longitudinal and transverse waves propagating in the elastic medium. As an illustration, we calculate the elastic scattering cross sections of a generic optical nanoantenna---a gold spherical nanoparticle---and discuss in detail the resonances found in such a system. We show how these resonators act as efficient and selective elastic antennas for both longitudinal and transverse radiation from localized elastic emitters, substantially suppressing or enhancing their emission rates. The elastic modes of the resonator are further characterized by introducing physical parameters analogous to those used in electromagnetism, such as an elastic Purcell factor, or an effective mode volume.
While we limit our investigations in this Letter to a dipolar point force coupled to the displacement field for simplicity, we plan to extend our formalism to additional sources for consideration of complex systems in future work. Similar modifications of the optical Purcell effect have been introduced to describe the coupling of magnetic dipolar, or multipolar emitters, to electromagnetic waves.\cite{Schmidt12} This framework offers a direct path towards a quantum-mechanical description of the interaction in elastic nanosystems and can be readily applied to explore other phenomena in the \textit{weak-} and \textit{strong-elastic coupling} regimes.
	
\begin{figure}[htbp!]
	\begin{center}
		\includegraphics[width=\columnwidth]{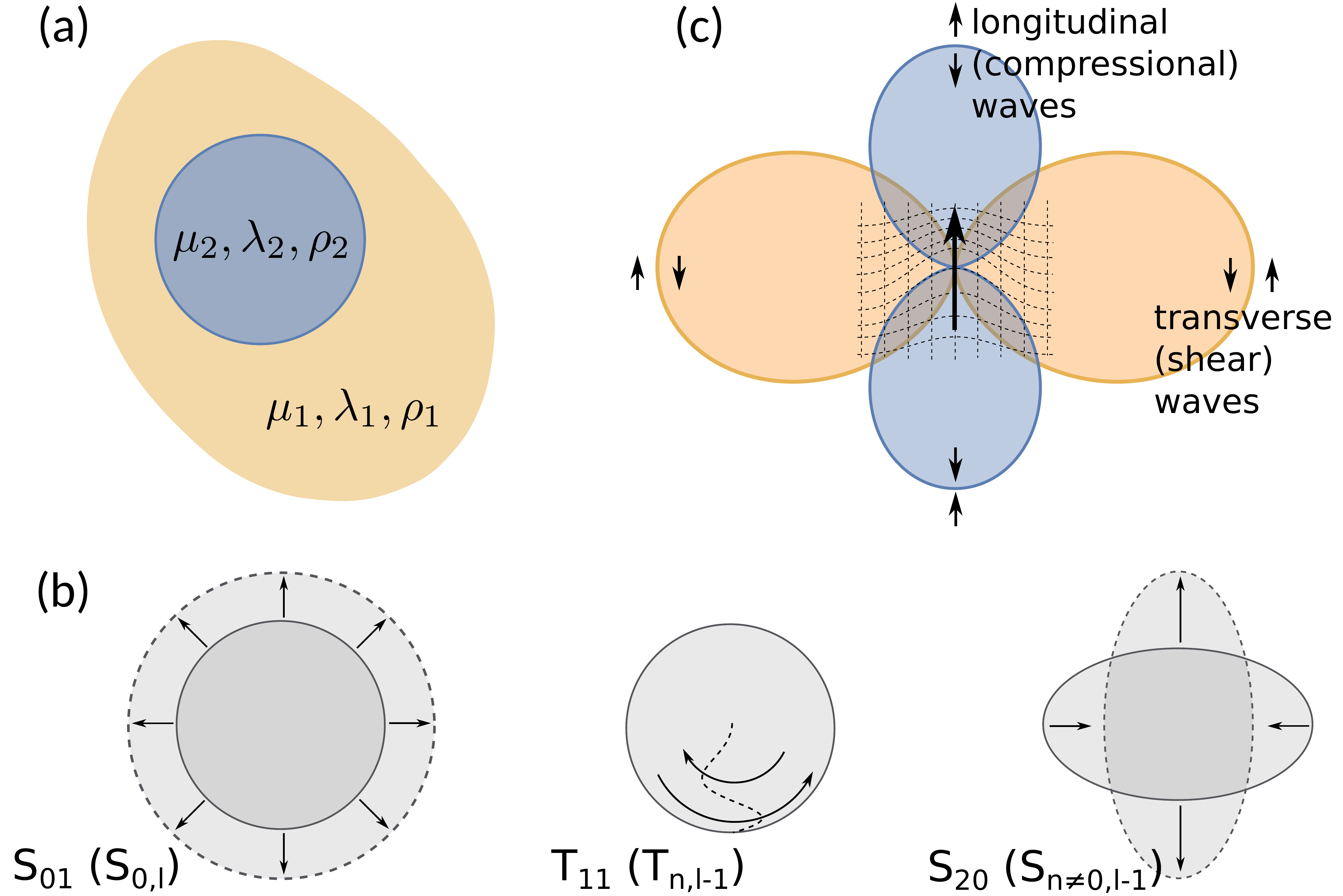}
		\caption{Schematics of (a) an elastic spherical nanoantenna, (b) its three families of m\textsl{}odes and (c) elastic localized emitters. (a) The antenna (medium) is characterized by its density $\rho_2$ ($\rho_1$) and two Lam\'e parameters: $\lambda_2$ ($\lambda_1$) and $\mu_2$ ($\mu_1$). (b) Elastic modes of spherical resonators (left to right): \textit{compressional} $S_{0,l}$, \textit{pure shear} $T_{n,l-1}$, and \textit{mixed} $S_{n\neq 0,l-1}$. (c) Radiation pattern from an elastic emitter (large black arrow) in an isotropic, homogeneous elastic medium. Transverse (shear) waves carry the energy in the plane perpendicular to the axis of the emitter, while the longitudinal (compressional) waves radiate along it. Small black arrows depict the orientation of displacement fields in the far-field, and the dashed grid illustrates the displacement field in the near-field.}
		\label{Fig1}
	\end{center}
\end{figure}
	
The generic elastic nanoantenna shown in Fig. \ref{Fig1}(a) is a homogeneous isotropic solid sphere characterized by Lam\'e parameters $\lambda_2$, $\mu_2$, and density $\rho_2$, and the velocities of longitudinal and transverse waves.\cite{auld1973acoustic} We consider the surrounding medium to be similarly isotropic, solid and homogeneous, with parameters $\lambda_1$, $\mu_1$, $\rho_1$, and supporting both transverse and longitudinal waves. To pursue elastic systems resonant at the typical mechanical frequencies of interest---a few GHz\cite{PhysRevX.5.031031,chu2017quantum,ruskov2013chip}---we can consider submicron metallic spheres.\cite{ahmed2017understanding,yu2013damping,pelton2011mechanical} Coincidentaly, such particles exhibit large \textit{optical} Purcell factors, and have therefore been explored as building blocks for optical nanodevices.\cite{Schmidt12,PhysRevLett.110.237401,Koenderink:10} Rescaling towards THz frequencies characteristic of Raman scattering \cite{LeRu20091} would likely be limited by the large viscosity of the composite materials in this regime.

To establish the formalism used throughout this work, let us consider the decomposition of the displacement field $\mathbf{u}$ into longitudinal ($\mathbf{u}_l$) and shear ($\mathbf{u}_s$) monochromatic waves propagating in the environment with Lam\'e parameters $\lambda_1$ and $\mu_1$ at velocities $v_{l,1}$ and $v_{s,1}$, respectively.\cite{ying1956scattering,einspruch1960scattering,pao1973diffraction,achenbach2012wave,graff2012wave} In the homogeneous isotropic medium, these displacement fields satisfy the wave equation $\omega^2 \pmb{\chi} + v_{\chi}^2 \nabla^2 \pmb{\chi} = 0$ for $(\pmb{\chi},v_{\chi})=(\mathbf{u}_s,v_{s,1})$ and $(\pmb{\chi},v_{\chi})=(\mathbf{u}_l,v_{l,1})$.
We expand the shear field into two families of vector spherical harmonics (VSHs):
\begin{equation}
\mathbf{M}_{mn} = \nabla \times (\mathbf{r} \Psi_{mn}), \quad \mathbf{N}_{mn} = \frac{v_s}{\omega}\nabla \times \mathbf{M}_{mn},
\end{equation}
defined by a scalar potential $\Psi_{mn}$ fulfilling the scalar wave equation with velocity $v_{s,1}$ ($(\chi,v_{\chi})=(\Psi_{mn},v_{s,1})$). As for EM, the transverse VSHs are constructed from the associated Legendre polynomials $P_n^m$ and spherical Bessel functions $j_n$, $y_n$, $h_n^{(1)}$ and $h_n^{(2)}$.\cite{bohren2008absorption} Similarly, the longitudinal waves propagating in the medium are described by one family of VSHs $\mathbf{L}_{mn} = \nabla \Phi_{mn}$ derived from the scalar solution $\Phi_{mn}$ to the wave equation with velocity $v_{l,1}$. 
Full expressions for all the VSHs can be found in Appendix A and in Refs. \citenum{ying1956scattering, einspruch1960scattering}.

These three families of VSHs can be used to represent any physical wave impinging on the scattering sphere, and simultaneously used to characterize the quasimodes of the vibrating sphere, shown in Fig. \ref{Fig1}(b). \textit{Compressional} modes, denoted usually as $S_{0,l}$,\cite{Notation,auld1973acoustic} which include the lowest-order \textit{breathing mode} (Fig. \ref{Fig1}(b), left panel), are given by the $\mathbf{L}_{mn}$ VSHs; \textit{pure shear} (or \textit{torsional}) modes $T_{n,l-1}$ with vanishing radial displacement fields are given by the $\mathbf{M}_{mn}$ VSHs, and include the lowest-order $T_{11}$ mode depicted in the middle panel in Fig. \ref{Fig1}(b)); finally, \textit{mixed} modes $S_{n\neq 0,l-1}$ include the lower-order mode shown in the right panel in Fig. \ref{Fig1}(b), in which compression along one axis induces expansion in the normal plane. The modes of this last family are described by linear combinations of the $\mathbf{N}_{mn}$ and $\mathbf{L}_{mn}$ harmonics. {
Note that while subscript $n$ corresponds to the radial dependence of both the modes of the sphere and the VSHs, subscript $l$ indexes the resonant frequencies for a given $n$ and is, in general not related to the azimuthal number $m$ of the VSH.\cite{auld1973acoustic}} 

Since we identify three families of elastic VSHs, in contrast to two found in the electromagnetic Mie theory, we can expect that the elastic boundary conditions (BCs) between the medium and the sphere will be more involved. Indeed, they require that all components of the displacement field $\mathbf{u}$, as well as normal components of the stress field $\mathbf{T}\cdot \mathbf{\hat{n}}$ be continuous. 
For the sphere centered at the origin of the coordinate system, these normal components correspond to the radial elements of the stress tensor $\mathbf{T}\cdot \mathbf{\hat{r}} = (T_{rr}, T_{r\theta}, T_{r\phi})$: \begin{equation}\label{stress.def.text1}
\frac{T_{rr}}{2 \rho v_s^2}=\frac{\sigma \nabla\cdot \mathbf{u}}{1-2\sigma} + u_{r,r}, \quad
\frac{T_{r\theta}}{\mu} = u_{\theta,r} - \frac{u_\theta}{r}+ \frac{u_{r,\theta}}{r},
\end{equation}
\begin{equation}\label{stress.def.text2}
\frac{T_{r\phi}}{\mu} =\frac{u_{r,\phi}}{r \sin(\theta)} + u_{\phi,r}- \frac{u_\phi}{r},
\end{equation}
where $u_{i,j}=\partial_j u_i$, and Poisson's ratio $\sigma$ is related to the Lam\'e parameters through $\sigma=\lambda/[2(\lambda+\mu)]$.\cite{auld1973acoustic} By numerically solving these BCs in the truncated basis of VSHs (see Appendix B, and Refs. \citenum{ying1956scattering,einspruch1960scattering}), we identify eigenmodes of the system characterized by complex eigenfrequencies. Following the literature on cavity QED, we dub these \textit{quasinormal modes} (QNMs).\cite{PhysRevLett.110.237401} The QNMs identified in the nanoantenna system include pure shear, mixed and compressional modes described earlier.
 	
\begin{figure}[htbp!]
	\begin{center}
		\includegraphics[width=\columnwidth]{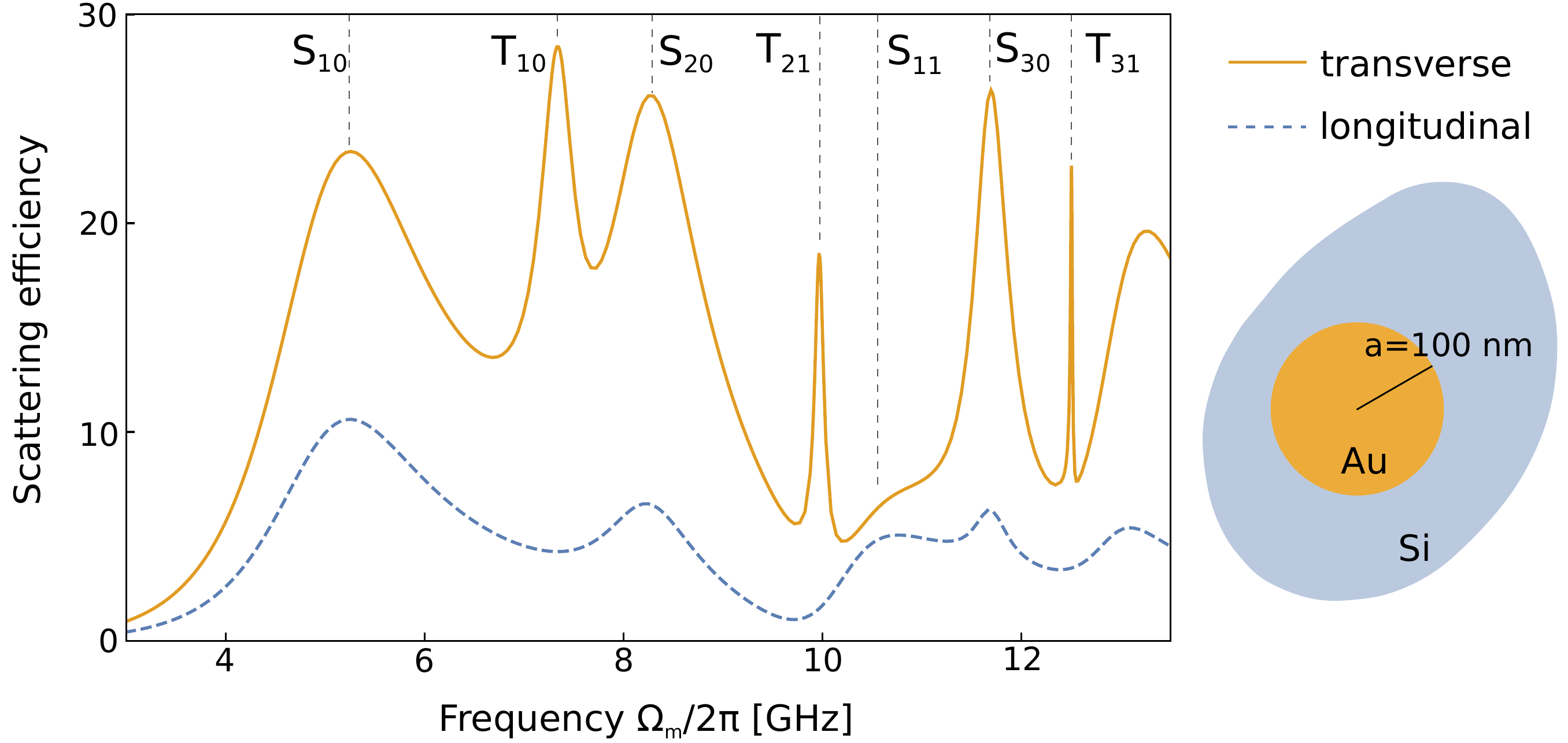}
		\caption{Scattering efficiencies of elastic longitudinal (blue dashed lines) and transverse (orange solid lines) plane waves by a spherical submicron radius gold resonator immersed in silicon in the lowest-order resonance region.\cite{weber2002handbook} 
		}
		\label{Fig2}		
	\end{center}
\end{figure}

In Fig.~\ref{Fig2} we present a comparison of the scattering efficiency (defined as the scattering cross section normalized by the geometric cross section of the sphere; see Appendix B3) of transverse and longitudinal plane wave excitations for a spherical $100~$nm radius gold nanoantenna ($\lambda_{\textrm{Au}}=154$~GPa, $\mu_{\textrm{Au}}=30$~GPa, $\rho_{\textrm{Au}}=19.3$~g/cm$^3$) immersed in silicon ($\lambda_{\textrm{Si}}=52$~GPa, $\mu_{\textrm{Si}}=66$~GPa, $\rho_{\textrm{Si}}=2.3$~g/cm$^3$). 
The lowest energy mixed ($S_{10}$, $S_{20}$, $S_{11}$, $S_{30}$) and pure shear modes ($T_{10}$, $T_{21}$, $T_{31}$) are marked with dashed lines. Shear QNMs, resonant at $\omega_{T_{10}}/2\pi \approx 7.34$~GHz, $\omega_{T_{21}}/2\pi \approx 9.97$~GHz, $\omega_{T_{31}}/2\pi \approx 12.5$~GHz, are characterized by quality factors from $18$ to $550$. Since these modes are accessible only to transverse plane waves, they would not be easily excited if the medium supported solely longitudinal, pressure-like waves. Compressional modes are found for larger frequencies, with $\omega_{S_{01}}/2\pi \approx 12.8$~MHz and $Q_{S_{01}} \approx 12$.

Having analyzed
the response of {generic} elastic nanoantennas, we can consider their application for controlling the dynamics of nearby phononic emitters. A number of such elastic sources have been recently reported, exhibiting both bosonic\cite{merklein2015enhancing,vahala2009phonon} and fermionic\cite{chu2017quantum,ruskov2013chip,Hong203} statistics of the emitted phonons, and are being investigated for applications in phonon-based quantum networks. 

The choice of model for an elastic emitter, and its coupling to the mechanical vibration of the environment, depends on the particular physical realization of the source. In this work we consider the simplest scenario, and model the elastic emitter as an oscillating, localized linear force $\mathbf{f}e^{-i\Omega t}$. This form of an elastic source can be realized in various optomechanical systems, for example by considering magnetostrictive inclusions in the structure,\cite{forstner2012cavity} or an absorbing nanoparticle vibrating due to harmonic radiation pressure.\cite{jain2016direct} We note that such a canonical optomechanical setup has been realized in numerous studies \cite{RevModPhys.86.1391} and that the elastic radiation rate has proven to be the critical factor in reaching interesting regimes of cooling of the mechanical degree of freedom as well as phonon lasing.

For such an emitter embedded in a homogeneous, lossless medium, we calculate the power of the emitted elastic radiation from the work done by the force on the displacement field $\mathbf{u}(\mathbf{r}_m)e^{-i\Omega t}$ at the position of the emitter ($\mathbf{r}_m$):
\begin{equation}\label{powerP0}
P_0 = \frac{d W}{dt} = \frac{\Omega_m}{2}\Im[\mathbf{u}(\mathbf{r}_m) \cdot \mathbf{f}^*].
\end{equation}
This radiation will dissipate the energy of the emitter with rate $\gamma_0$.\cite{novotny2012principles} The displacement $\mathbf{u}(\mathbf{r}_m)$ can be found by applying the definition of the elastic
Green function {$\mathbf{u}(\mathbf{r}) = \mathbf{G}(\mathbf{r},\mathbf{r}_m) \mathbf{f}(\mathbf{r}_m)$}. In the environment characterized by $(\lambda_1,\mu_1,\rho_1)$, in terms of ${\mathbf{d}=\mathbf{r}-\mathbf{r}_m}$, the dyadic Green function is\cite{snieder,ben2012seismic}
\begin{align}\label{Green.function}
\mathbf{G}_0(\mathbf{d}) = &\frac{i k_{l1}}{12 \pi (\lambda_1 + 2\mu_1)}\left[\mathbbm{1} h_0^{(1)}(k_{l1} d)+ \left(\mathbbm{1}-3\hat{\mathbf{d}}\hat{\mathbf{d}}^T\right) h_2^{(1)}(k_{l1} d)\right] \\ \nonumber
&-\frac{i k_{s1}}{12 \pi \mu_1}\left[-2\mathbbm{1} h_0^{(1)}(k_{s1} d)+ \left(\mathbbm{1}-3\hat{\mathbf{d}}\hat{\mathbf{d}}^T\right) h_2^{(1)}(k_{s1} d)\right],
\end{align} 
where $\mathbbm{1}$ is the identity tensor, $h_0^{(1)}$ and $h_2^{(1)}$ are the spherical Hankel functions of the first kind, of the 0th and 2nd order. Here, $k_{l1} = \Omega/v_{l1}$, and $k_{s1} = \Omega/v_{s1}$ are the wave vectors associated with longitudinal and transverse (shear) waves, respectively.

The second part of the Green function, describing the shear waves, has a form {similar} to that of the electromagnetic Green function, ensuring that the shear waves propagate in a dipole-like pattern, with the largest emission in the plane perpendicular to the axis of the force ($\theta=0$). The longitudinal waves have an identical near- and intermediate-field distribution (except for the wave vector), but the far field has a $\cos^2 \theta$-like distribution (see Fig. \ref{Fig1}(c) and Appendix A).

The radiated power $P_0$ in the homogeneous medium can be calculated from the Green function (see Appendix C) as 
\begin{equation}
P_0 = \frac{\Omega |\mathbf{f}|^2}{12 \pi}\left[\frac{k_{s1}}{\mu} + \frac{k_{l1}}{2(\lambda_1 + 2\mu_1)}\right].
\end{equation} 
{Unlike for the power emitted by an electromagnetic dipolar emitter, the elastic $P_0$ exhibits quadratic, rather than quartic, dependence on the emission frequency $\Omega$. This is due to the difference in the dispersion of the inhomogeneous term in the wave equations for electric field $\mathbf{E}$ compared to displacement $\mathbf{u}$.\cite{auld1973acoustic} 

The radiation of an emitter in an arbitrary system can be calculated by expanding Eq.~(\ref{powerP0}), including the Green function that represents scattering on an elastic antenna $\mathbf{G} = \mathbf{G}_0 + \mathbf{G}_S$, and writing the enhancement as
\begin{align}\label{power.enh}
\frac{P}{P_0} &= 1 + \frac{6 \pi}{|\mathbf{f}|^2} \frac{\Im\{\mathbf{f}^* \cdot  \mathbf{G}_S(\mathbf{r}_m,\mathbf{r}_m) \cdot \mathbf{f}\}}{\frac{k_{l1}}{2(\lambda_{1} + 2\mu_{1})} + \frac{k_{s1}}{\mu_{1}}}.
\end{align}
Consequently, the rate of energy dissipation $\gamma$ from the elastic emitter ${\mathbf{f}}$ near the antenna will be enhanced, or quenched with respect to the free-space rate $\gamma_0$ following $\gamma/\gamma_0 = P/P_0$. 
We note that alternate implementations of an elastic emitter necessarily call for a modification of the expression for the emitter-antenna interaction of Eq.~(\ref{powerP0}), and the radiation power of Eq.~(\ref{power.enh}). For example, spatially localized Brillouin scattering-induced lateral vibrations of a waveguide,\cite{merklein2017chip} or orbital transitions in SiV defects in diamonds, would need to be modeled as quadrupolar elastic sources coupled to a strain tensor,\cite{ruskov2013chip} rather than displacement field. Nevertheless, (an appropriately modified) Eq.~(\ref{power.enh}) can be used to extract the total LDES from the trace of the complete Green function $\Tr \{\Im[\mathbf{G}(\mathbf{r}_m,\mathbf{r}_m)]\}$.\cite{novotny2012principles}

As an example, let us consider the decay rate enhancement $\gamma/\gamma_0$ of a dipolar emitter positioned $z=1.2a$ from the center of the previously analyzed spherical gold scatterer embedded in silicon, shown schematically in Fig. 3(a). {By tracing the spectral positions of the features in the spectrum corresponding to the emitter oriented radially, we can identify the enhancement and suppression of the emission as originating from the coupling with the low-\textit{Q} mixed QNMs $S_{nl}$\cite{Notation}. Spectra for the azimuthally oriented emitter indicate additional, significant enhancements due to the coupling with higher-\textit{Q} shear QNMs $\left(T_{nl}\right)$.} {These assertions regarding coupling to the QNMs originate solely from the identification of the peak frequencies, since the underlying Eq.~(\ref{power.enh}) does not readily provide insight into the modal structure of the enhancement spectra.}

{It is thus instructive to consider a modal picture of the coupling between a localized emitter and the QNMs of an elastic nanoresonator. Such a picture completes the analogy with the optical Purcell factor,\cite{Koenderink:10,PhysRevB.91.195422,PhysRevLett.110.237401,esteban2014strong,mertens2007plasmon} originally introduced through the formalism of quantum electrodynamics, and enables consideration of nonclassical phonon emitters such as piezoelectric microantennas coupled to superconducting qubits,\cite{chu2017quantum} photonic crystals in the single-phonon regime,\cite{Hong203} or GHz transitions in color centers in diamond.\cite{PhysRevB.94.214115}} To this end, we seek the elastic \textit{Purcell factor} $F_\alpha$ which characterizes a single mode $\alpha$ of the cavity. As Koenderink\cite{Koenderink:10} pointed out while analyzing plasmonic cavities, the Purcell factor provides a reasonable approximation of the rate enhancement only if (i) the normal modes of the system can be identified and (ii) one of these modes is dominant. Since we have already found the QNMs of the system, and recognized cases where the high-\textit{Q}, well-separated shear modes dominate the response, we adopt this framework. However, since the Purcell factor is traditionally expressed solely in terms of the characteristics of a single cavity mode, assuming optimal positioning or orientation of the emitter, the Purcell factor should be treated as an upper bound for the spontaneous rate enhancement factor $\gamma/\gamma_0$. We address this limitation later.

We introduce the expansion of the nanoresonator displacement field into quantized QNMs:\cite{sipe2016hamiltonian} 
$\hat{\mathbf{u}}(\mathbf{r}) = \sum_{\alpha} \hat{\mathbf{u}}_\alpha(\mathbf{r}) = \sum_{\alpha}\sqrt{{\hbar}/(2 \Omega_\alpha \rho(\mathbf{r}))} \left[\hat{b}_\alpha {\mathbf{U}_\alpha(\mathbf{r})}+H.c.\right],$
where the displacement field is represented as a sum over modes $\alpha$ with resonant frequencies $\Omega_\alpha$, decay rates $\kappa_\alpha$, normalized mode profiles $\mathbf{U}_\alpha$, and phonon creation (annihilation) operators $\hat{b}_\alpha^\dag$ ($\hat{b}_\alpha$).
An interaction of the modes with the quantum emitter positioned at $\mathbf{r}_m$, described as a two-level system $\hat{\mathbf{f}} = \tilde{\mathbf{f}} (\ket{g}\bra{e} + \ket{e}\bra{g})$, is expressed by a Hamiltonian ${\hat{H}_I = \hat{\mathbf{f}}\cdot \hat{\mathbf{u}}(\mathbf{r}_m)}$. {Formally, the elastic Purcell factor is then defined as the ratio of the upper bound for} the rate of spontaneous emission $\gamma_\alpha$ of the emitter excitation into the cavity mode $\alpha$, and 
$\gamma_0$:\cite{esteban2014strong,auffeves2010controlling}
\begin{align}\label{Purcell.def}
F_\alpha &= \frac{\gamma_\alpha}{\gamma_0} = \frac{6\pi}{\rho_2 \rho_1^{1/2}\phi_1} \frac{Q_\alpha}{\Omega_\alpha^3 V_{\alpha,\text{eff}}},
\end{align}
where $\screen_i = \mu_i^{-3/2} +\frac{1}{2} \left(2\mu_i+\lambda_i\right)^{-3/2}$ can be interpreted as a screening factor, the quality factor of the resonator $Q_{\alpha}$ is given by $\Omega_\alpha/\kappa_\alpha$, and the mode displacement field defines the effective mode volume $V_{\text{eff}}$:
\begin{equation}\label{Veff.def}
	V_{\text{eff},\alpha} = \frac{\int d\mathbf{r} \rho(\mathbf{r}) |\mathbf{u}_\alpha(\mathbf{r})|^2 }{\text{max}[\rho(\mathbf{r})|\mathbf{u}_\alpha(\mathbf{r})|^2]}.
\end{equation}
Interestingly, $F_\alpha$ has a form very similar to its electromagnetic counterpart, even exhibiting $\Omega_\alpha^{-3}$ dependence despite the marked dispersion differences of the free-space radiation power $P_0$ pointed out earlier. A detailed derivation of $F_\alpha$ and $V_{\text{eff},\alpha}$ is given in Appendix D. We note that an equivalent expression for $V_{\text{eff},\alpha}$ was derived by Eichenfield \textit{et al.}\cite{eichenfield2009optomechanical} in an alternative way, by analyzing the energy density in optomechanical crystals. 

\begin{figure}[htbp!]
	\begin{center}
		\includegraphics[width=\columnwidth]{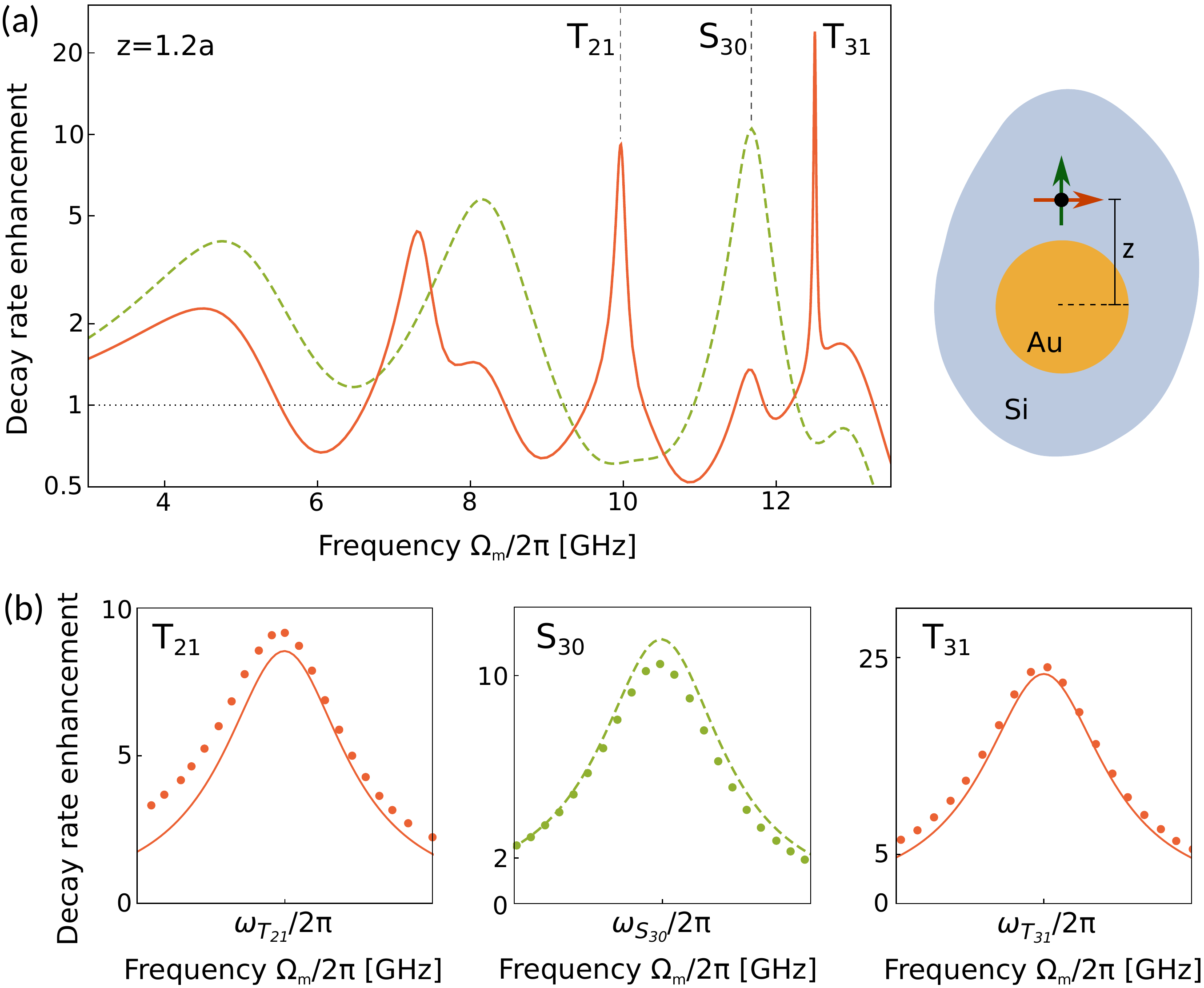}
		\caption{Decay rate enhancement for an elastic emitter near a nanosphere. (a) Spectrum of decay rate enhancement $\Gamma/\Gamma_0$ (Eq.~(\ref{power.enh})) for the emitter oriented radially (green dashed line) or azimuthally (red solid line). (b) Magnified regions around three modes: $T_{21}$, $S_{30}$, and $T_{31}$ coupled to by an azimuthally ($T_{21}$ and $T_{31}$) or radially ($S_{30}$) polarized emitter. Points represent the exact calculations (Eq.~(\ref{power.enh})), while the solid lines describe coupling to the quasinormal modes of the resonator (Eq.~(\ref{corr.Pf2})). 
		}
	\end{center}
\end{figure}

We can thus calculate the characteristics of a few lowest-order modes of gold scatterers in silicon, marked in Fig.~3(a) with dashed lines. For the pure shear $T_{21}$ mode centered at $\Omega_{T_{21}}/2\pi \approx 9.97$~GHz, the effective mode volume $V_{\textrm{eff},T_{21}}/V_{\textrm{geom}} \approx 0.25$ and the quality factor $Q_{T_{21}} \approx 90$ yield the Purcell factor $F_{T_{21}} \approx 109$. Similarly, for the $T_{31}$ mode the parameters are $\Omega_{T_{31}} /2\pi \approx 12.5$~GHz, $V_{\textrm{eff},T_{31}}/V_{\textrm{geom}} \approx 0.17$, $Q_{T_{31}} \approx 550$, and $F_{T_{31}} \approx 480$. Finally, for the mixed mode $S_{30}$, we find $\Omega_{S_{30}}/2\pi \approx 11.69$~GHz, $V_{\textrm{eff},S_{30}}/V_{\textrm{geom}} \approx 0.17$, $Q_{S_{30}} \approx 28$, and $F_{T_{30}} \approx 32$. These results indicate that the considerable Purcell factors stem, as expected for the low-order Mie resonances,\cite{Schmidt12} from high $Q_\alpha$'s of the QNMs rather than reduced effective mode volumes $V_{\text{eff},\alpha}$.

We note that while neither silicon nor gold are isotropic materials\cite{saviot2009acoustic} (characterized by elastic anisotropic ratios A$_\text{Si}\approx 1.6$ and A$_\text{Au}\approx 0.6$), their simplified description does not incur significant shifts of resonant frequencies, nor changes in Purcell factors. Additionally, we have verified by full numerical calculations that when viscous damping in gold is included through the complex Young's modulus,\cite{doi:10.1021/acs.nanolett.8b00559} the quality factors and the Purcell factors characterizing higher-order elastic modes are significantly quenched. While further studies of viscous damping are required to understand its exact magnitude, this mechanism will likely promote the use of lowest-order elastic modes in nanostructures for enhancing radiative emission from localized emitters.

As we have noted earlier, the Purcell factors are upper bounds for the decay rate enhancements, and so the estimated values of $F_\alpha$ are significantly larger than those calculated with Eq.~(\ref{power.enh}) and shown in Fig.~3(a). This difference is mostly due to the fact that, much like for dielectric high-refractive index electromagnetic Mie scatterers,\cite{Schmidt12} the displacement field in the investigated cases is localized in the \textit{slow} medium, i.e. gold, where the longitudinal and shear waves propagate $v_{l1}/v_{l2}\approx 2.7$ and $v_{s1}/v_{s2}\approx 4.3$ times slower than in the environment, respectively. Achieving such ratios of velocities between the medium and the antenna is difficult in electromagnetic systems.

To improve on the accuracy of this approach, it is instructive to modify the definition of the Purcell factor, introducing the dependence on the polarization, position, and the detuning of the emitter with respect to the cavity resonance $\Omega_\alpha$ (see Appendix D):
\begin{align}\label{corr.Pf2}
\tilde{F}_{\alpha}(\mathbf{r}_m) = F_{\alpha} \frac{ \Omega_\alpha}{\Omega_m} \frac{|\tilde{\mathbf{f}} \cdot \mathbf{u}_\alpha(\mathbf{r}_m)|^2}{ |\tilde{\mathbf{f}}|^2\max_\mathbf{r}|\mathbf{u}_\alpha(\mathbf{r})|^2 }\frac{(\kappa_\alpha/2)^2}{(\Omega_\alpha- \Omega_m)^2+(\kappa_\alpha/2)^2}.
\end{align}

Functions $\tilde{F}_{\alpha}(\mathbf{r}_m)$, calculated for the shear modes and mixed modes of the gold scatterer, are shown in the three panels of Fig.~3(b) with lines and correspond well to the exact results obtained from Eq.~(\ref{power.enh}) depicted with points. The discrepancies found for the two lower \textit{Q} modes $T_{21}$ and $S_{30}$ are due to the additional coupling of the emitter with other QNMs and the continuum of free-space radiation modes.

In conclusion, we have introduced the formalism inherited from quantum electrodynamics to characterize elastic nanoresonators, and found that such structures support modes with large Purcell factors determined by considerable $Q$s, rather than significantly reduced effective mode volumes. In particular, submicron gold particles, widely used in nanooptics, can readily enhance of spontaneous emission from phononic GHz emitters. Furthermore, such elastic nanoantennas offer robust spatial and spectral control over the dissipation of mechanical vibrations, and can be therefore used to introduce subwavelength control of phonon dynamics to various mechanical \cite{merklein2017chip,behunin2017engineering,maldovan2013sound} and optomechanical systems.\cite{eichenfield2009optomechanical,vahala2009phonon}

We acknowledge support from the Australian Research Council (ARC) Discovery Project No. DP160101691 and ARC Center of Excellence CUDOS (Grant No. CE110001018).

\textit{Note added} After the paper was accepted for publication, recent related contributions were brought to our attention \cite{doi:10.1121/1.4990010,PhysRevLett.120.114301}.

\bibliographystyle{apsrev4-1} 
\bibliography{bibliography}

\clearpage
\onecolumngrid
\appendix

%
%

\section{Numerical solution of the scattering problem}\label{Expansion}

\subsubsection{Expansion into longitudinal waves}
Longitudinal fields are expanded by a set of longitudinal VSHs:
\begin{align}
\mathbf{L}_{omn} = \nabla \Phi_{omn} &= \hat{\mathbf{r}}\frac{\mathrm{d} z_n(k_l
	r)}{\mathrm{d} (k_l r)} P_n^{|m|}(\cos \theta ) \sin m\phi  + \hat{\mathbf{\theta}}\frac{z_n(k_l
	r) }{k_lr} \frac{\mathrm{d} P_n^{|m|}(\cos \theta )}{\mathrm{d} \theta} \sin m\phi  \\ \nonumber
&+ \hat{\mathbf{\phi}}\frac{z_n(k_l
	r)}{k_lr}\frac{P_n^{|m|}(\cos \theta )}{\sin \theta}m \cos m\phi,
\end{align}
\begin{align}
\mathbf{L}_{emn} = \nabla \Phi_{emn} &= \hat{\mathbf{r}}\frac{\mathrm{d} z_n(k_l
	r)}{\mathrm{d} (k_l r)} P_n^{|m|}(\cos \theta ) \cos m\phi + \hat{\mathbf{\theta}}\frac{z_n(k_l
	r) }{k_lr} \frac{\mathrm{d} P_n^{|m|}(\cos \theta )}{\mathrm{d} \theta} \cos m\phi \\ \nonumber
&- \hat{\mathbf{\phi}}\frac{z_n(k_l
	r)}{k_lr}\frac{P_n^{|m|}(\cos \theta )}{\sin \theta}m \sin m\phi ,
\end{align}
derived from the scalar potentials
\begin{equation}
\Phi_{omn} = z_n(k_l r) P_n^m (\cos \theta) \sin m\phi, \quad 	\Phi_{emn} = z_n(k_l r) P_n^m (\cos \theta) \cos m\phi.
\end{equation}
Radial function $z_n$ is any of the spherical Bessel functions of the first kind $j_n$, the second kind $y_n$, or their linear combinations dubbed as spherical Hankel functions $h_n^{(1)}$.\cite{bohren2008absorption} We will choose them following the derivation of the electromagnetic Mie theory, adding a superscript $(1)$ to the VSHs which use $j_n$, and make up fields necessarily finite at the origin of coordinates. Conversely, the superscript $(3)$ will denote the VSHs derived from the generating functions with $h_n^{(1)}$.

\subsubsection{Expansion into transverse waves}
The transverse fields should be expanded into the set of VSHs familiar from the electromagnetic Mie theory:
\begin{align}
\mathbf{M}_{emn} = \nabla \times (\mathbf{r} \Psi_{emn}) = & -\hat{\mathbf{\theta}} m\sin m\phi \frac{P_n^{|m|}(\cos \theta )}{\sin \theta} z_n(k_s r) - \hat{\mathbf{\phi}} \cos m\phi\frac{\mathrm{d} P_n^{|m|}(\cos \theta )}{\mathrm{d} \theta} z_n(k_s r),
\end{align}
\begin{align}
\mathbf{M}_{omn} = \nabla \times (\mathbf{r} \Psi_{omn}) = & \hat{\mathbf{\theta}} m\cos m\phi\frac{P_n^{|m|}(\cos \theta )}{\sin \theta} z_n(k_s r) - \hat{\mathbf{\phi}} \sin m\phi \frac{\mathrm{d} P_n^{|m|}(\cos \theta )}{\mathrm{d} \theta} z_n(k_s r),
\end{align}
\begin{align}
\mathbf{N}_{emn} = \frac{v_s}{\Omega} \nabla \times \mathbf{M}_{emn} = & \hat{\mathbf{r}} \cos m\phi n (n+1)P_n^{|m|}(\cos \theta ) \frac{z_n(k_s r)}{k_s r} +\hat{\mathbf{\theta}} \cos m\phi\frac{d P_n^{|m|}(\cos \theta )}{d \theta} \frac{1}{k_s r}\frac{\mathrm{d}}{\mathrm{d} r}[r z_n(k_s r)] \\ \nonumber
&- \hat{\mathbf{\phi}} m \sin m\phi \frac{ P_n^{|m|}(\cos \theta )}{\sin \theta} \frac{1}{k_s r}\frac{\mathrm{d}}{\mathrm{d} r}[r z_n(k_s r)],
\end{align}
\begin{align}
\mathbf{N}_{omn} = \frac{v_s}{\Omega} \nabla \times \mathbf{M}_{omn} = & \hat{\mathbf{r}} \sin m\phi n (n+1)P_n^{|m|}(\cos \theta ) \frac{z_n(k_s r)}{k_s r} +\hat{\mathbf{\theta}} \sin m\phi \frac{\mathrm{d} P_n^{|m|}(\cos \theta )}{\mathrm{d} \theta} \frac{1}{k_s r}\frac{\mathrm{d}}{\mathrm{d}r}[r z_n(k_s r)] \\ \nonumber
& + \hat{\mathbf{\phi}} m \cos m\phi\frac{ P_n^{|m|}(\cos \theta )}{\sin \theta} \frac{1}{k_s r}\frac{\mathrm{d}}{\mathrm{d} r}[r z_n(k_s r)],
\end{align}
derived from the scalar potentials
\begin{equation}
\Psi_{emn} = z_n(k_s r) P_n^m (\cos \theta) \sin m\phi \quad 	\Psi_{emn} = z_n(k_s r) P_n^m (\cos \theta) \cos m\phi.
\end{equation}
We note that the derivatives over function $z_n$ can be rewritten as in Ref. [\citenum{bohren2008absorption}]
\begin{equation}
\frac{d}{d r}[r z_n(k r)] = \frac{\mathrm{d}}{\mathrm{d} (k r)}\frac{\mathrm{d} (k r)}{\mathrm{d} r}[r z_n(k r)] = \frac{\mathrm{d}}{\mathrm{d} (k r)}[k r z_n(k r)].
\end{equation}

\section{Scattering on a sphere}
We will now consider the problem of scattering of the incident illumination, represented as a sum of these vector spherical harmonics, on a sphere with radius $|\mathbf{r}|=a$, centered at the origin of coordinate system. 

This particular problem, as well as the related questions regarding the response of a cavity, or a rigid body---both of which require different boundary conditions---were discussed in detail in the original contributions from Ying et al.\cite{ying1956scattering} and Einspruch et al.\cite{einspruch1960scattering}. 

\subsubsection{Fields inside and outside the sphere}
Following the notation convention used widely in the electromagnetic Mie theory, we denote the fields and stresses inside of the sphere by subscript $1$, the fields and stresses associated with the incident illumination by subscript $i$, and those related to the scattered fields by $s$. The sphere is made of material with $\rho_2$, $\lambda_2$ and $\mu_2$, while the embedding medium is made of material with $\rho_1$, $\lambda_1$ and $\mu_1$, as depicted in Fig. 1.

The displacement fields are represented as a sum of the longitudinal and transverse VSHs:
\begin{equation}\label{inc}
\mathbf{u}_i = \sum_{j=o,e} \sum_{n=1}^{\infty} \sum_{m=n}^{\infty} \left[a_{jmn}\mathbf{L}_{jmn}^{(1)}(k_{1l}) + b_{jmn}\mathbf{N}_{jmn}^{(1)}(k_{1s}) + c_{jmn}\mathbf{M}_{jmn}^{(1)}(k_{1s})\right],
\end{equation}
\begin{equation}\label{scatt}
\mathbf{u}_s = \sum_{j=o,e} \sum_{n=1}^{\infty} \sum_{m=n}^{\infty} \left[d_{jmn}\mathbf{L}_{jmn}^{(3)}(k_{1l}) + e_{jmn}\mathbf{N}_{jmn}^{(3)}(k_{1s}) + f_{jmn}\mathbf{M}_{jmn}^{(3)}(k_{1s})\right],
\end{equation}
\begin{equation}\label{trans}
\mathbf{u}_1 = \sum_{j=o,e} \sum_{n=1}^{\infty} \sum_{m=n}^{\infty} \left[g_{jmn}\mathbf{L}_{jmn}^{(1)}(k_{2l}) + h_{jmn}\mathbf{N}_{jmn}^{(1)}(k_{2s}) + i_{jmn}\mathbf{M}_{jmn}^{(1)}(k_{2s})\right].
\end{equation}
Here we have introduced a simplified notation where the only arguments of the VSHs are the wavenumbers governing the radial dependence, e.g. $k_{1l}$ denotes wavenumber associated with longitudinal wave propagating in the 1st medium (environment). Unless otherwise stated, the parameters of the VSH should be $\theta$, $\phi$ and $r=a$ (point on the surface of the scattering sphere).

\subsection{On notation - scalar product and orthogonality}
We will now introduce the scalar product of the two functions (e.g. the VSHs) defined as
\begin{equation}
\langle \mathbf{f}, \mathbf{g} \rangle = \int_0^{2\pi} \mathrm{d}\phi\int_0^{\pi} \mathrm{d}\theta~\mathbf{f}^*(a,\theta,\phi)\cdot \mathbf{g}(a,\theta,\phi)
\end{equation}
i.e. with the integration carried out on the surface of the scattering sphere.

We know that the transverse VSHs are orthogonal, but not orthonormal (in the sense of the scalar product defined above). This orthogonality does not extend to the longitudinal VSHs, as
\begin{equation}
\langle \mathbf{L}_{omn}, \mathbf{N}_{omn}\rangle\neq 0, \quad \langle \mathbf{L}_{emn}, \mathbf{N}_{emn}\rangle\neq 0,
\end{equation}
although 
\begin{equation}
\langle \mathbf{L}_{omn}, \mathbf{M}_{omn}\rangle = \langle \mathbf{L}_{emn}, \mathbf{M}_{emn}\rangle = 0.
\end{equation}

\subsection{Boundary conditions}\label{bcs}

We impose two conditions on the surface of the sphere: continuity of the displacement, and continuity of the normal component of the stress:
\begin{equation}
\mathbf{u}_i + \mathbf{u}_s = \mathbf{u}_1,
\end{equation}
\begin{equation}
\hat{\mathbf{r}} \cdot \mathbf{T}_i + \hat{\mathbf{r}} \cdot \mathbf{T}_s = \hat{\mathbf{r}} \cdot \mathbf{T}_1.
\end{equation}
The radial components of the stress field in isotropic media, in spherical coordinates, are listed in Eqs.~(2) and (3).

In order to formulate these BCs in terms of the algebraic equations, we will use the expansion of the incident, scattered and internal displacement fields given in Eqs.~(\ref{inc}-\ref{trans}), and the associated stresses, and derive equations for the coefficients $d_{jmn}$ to $i_{jmn}$ in terms of the coefficients $a_{jmn}$ to $c_{jmn}$. We will truncate the summation over $n$ index to span from 0 to $N$, and treat every value of $m$ independently. Note that the coefficients describing the incident field ($a_{jmn}$ to $c_{jmn}$) can be found from directly expanding the incident field and accounting for the non-orthonormality of the VSHs. We will discuss this derivation in more detail later. For each value of index $m$, we thus have to find coefficients for parameters $d$ through $i$, with indices $j=o,e$ and $n=0,1,...,N$. These parameters make up a set of $12(N+1)$ coefficients, and thus we will need $12(N+1)$ linear equations to solve them. 


\subsubsection{Representing the first BCs for the fields decomposed into VSHs}
We can rewrite the 1st boundary condition by moving the unknown elements of the displacement fields to one side:
\begin{equation}\label{inout}
\mathbf{u}_1 - \mathbf{u}_s = \mathbf{u}_i,
\end{equation}
or, using the expansion from Eqs.(\ref{inc}), (\ref{scatt}) and (\ref{trans}): 
\begin{align}\label{summar}
&\sum_{j=o,e} \sum_{n=0}^{N} \left[g_{jmn}\mathbf{L}_{jmn}^{(1)}(k_{2l}) + h_{jmn}\mathbf{N}_{jmn}^{(1)}(k_{2s}) + i_{jmn}\mathbf{M}_{jmn}^{(1)}(k_{2s}) - d_{jmn}\mathbf{L}_{jmn}^{(3)}(k_{1l}) - e_{jmn}\mathbf{N}_{jmn}^{(3)}(k_{1s}) - f_{jmn}\mathbf{M}_{jmn}^{(3)}(k_{1s})\right]\\ \nonumber
&= \sum_{j=o,e} \sum_{n=1}^{N}  \left[a_{jmn}\mathbf{L}_{jmn}^{(1)}(k_{1l}) + b_{jmn}\mathbf{N}_{jmn}^{(1)}(k_{1s}) + c_{jmn}\mathbf{M}_{jmn}^{(1)}(k_{1s})\right],
\end{align}
for each $m$. We will be treating each $m$ separately, as due to the axial symmetry of the scatterer, BCs do not mix VSHs with different $m$'s. 

We now introduce an ordering into the numbering over indices, and assume that first we list all the coefficients for a fixed $n=0$, with $o$'s and then with $e$'s, then $n=1$ etc., so that the coefficients making up the unknown field $\mathbf{u}_{\textrm{un}} = \mathbf{u}_1 - \mathbf{u}_s$ field (LHS of Eq.~(\ref{inout})) form a column vector with $12(N+1)$ coefficients
\begin{equation}
\mathcal{C}_{\textrm{un}}^{m} = \left[\mathcal{C}_{\textrm{un}}^{m,n=0}, \mathcal{C}_{\textrm{un}}^{m,n=1}, ..., \mathcal{C}_{\textrm{un}}^{m,n=N}\right]^T,
\end{equation}
where each $\mathcal{C}_{\textrm{un}}^{m,n}$ denotes a list of coefficients
\begin{equation}
\mathcal{C}_{\textrm{un}}^{m,n} = \left[\underbrace{g_{omn},h_{omn},i_{omn},g_{emn},h_{emn},i_{emn}}_{\text{coefficients making up }\mathbf{u}_1},\underbrace{-d_{omn},-e_{omn},-f_{omn},-d_{emn},-e_{emn},-f_{emn}}_{\text{coefficients making up }\mathbf{u}_s}\right].
\end{equation}
The minus signs are there to reflect that the LHS of Eq.~(\ref{inout}) is a difference of the internal and scattered fields. We introduce a similar notation for the VSHs:
\begin{equation}
\mathcal{P}_{\textrm{un}}^m =
\left[\mathcal{P}_{\textrm{un}}^{m,n=0}, \mathcal{P}_{\textrm{un}}^{m,n=1}, ..., \mathcal{P}_{\textrm{un}}^{m,n=N}\right],
\end{equation}
where each $\mathcal{P}_{\textrm{un}}^{m,n}$ denotes a list of coefficients
\begin{align}
\mathcal{P}_{\textrm{un}}^{m,n} = \left[\right.&\mathbf{L}_{omn}^{(1)}(k_{2l}),\mathbf{N}_{omn}^{(1)}(k_{2s}),\mathbf{M}_{omn}^{(1)}(k_{2s}),\mathbf{L}_{emn}^{(1)}(k_{2l}),\mathbf{N}_{emn}^{(1)}(k_{2s}),\mathbf{M}_{emn}^{(1)}(k_{2s}),  \\ \nonumber 
& \left.\mathbf{L}_{omn}^{(3)}(k_{1l}),\mathbf{N}_{omn}^{(3)}(k_{1s}),\mathbf{M}_{omn}^{(3)}(k_{1s}),\mathbf{L}_{emn}^{(3)}(k_{1l}),\mathbf{N}_{emn}^{(3)}(k_{1s}),\mathbf{M}_{emn}^{(3)}(k_{1s})\right].
\end{align}
Note that $\mathcal{P}_{\textrm{un}}^m$ is now a vector of $12(N+1)$ elements, each one a vector field on a surface of the sphere with angular parameters $\theta$ and $\phi$. While this may seem like a tedious complication, it allows us to handily write down the LHS of Eq.~(\ref{inout}) simply as $(\mathcal{P}_{\textrm{un}}^m)^T \mathcal{C}_{\textrm{un}}^{m}$.

We can introduce analogous notation for the RHS of Eq.~(\ref{inout}), a column vector of $6(N+1)$ coefficients $\mathcal{C}_{i}^{m}$, and a vector of $6(N+1)$ VSHs $\mathcal{P}_{i}^{m}$, and get
\begin{equation}
( \mathcal{P}_{i}^m)^T \mathcal{C}_{i}^{m} = \mathbf{u}_i.
\end{equation}

The coefficients defining the incident field $\mathbf{u}_i$ ($a_{jmn}$, $b_{jmn}$ and $c_{jmn}$) are found by projecting this field on the surface of the sphere, and accounting for the non-orthonormality of the expansions functions, i.e. by solving the following set of equations:
\begin{align}
&\begin{bmatrix}
\langle \mathbf{L}_{om0}^{(1)}(k_{1l}),\mathbf{L}_{om0}^{(1)}(k_{1l}) \rangle & \langle \mathbf{L}_{om0}^{(1)}(k_{1l}),\mathbf{N}_{om0}^{(1)}(k_{1s}) \rangle & \dots & \langle \mathbf{L}_{om0}^{(1)}(k_{1l}),\mathbf{M}_{emN}^{(1)}(k_{1s})\\
\langle \mathbf{N}_{om0}^{(1)}(k_{1l}),\mathbf{L}_{om0}^{(1)}(k_{1l}) \rangle & \langle \mathbf{N}_{om0}^{(1)}(k_{1s}),\mathbf{N}_{om0}^{(1)}(k_{1s}) \rangle & \dots & \langle \mathbf{N}_{om0}^{(1)}(k_{1s}),\mathbf{M}_{emN}^{(1)}(k_{1s})\\
\hdotsfor{4} \\
\langle \mathbf{M}_{emN}^{(1)}(k_{1l}),\mathbf{L}_{om0}^{(1)}(k_{1l}) \rangle & \langle \mathbf{M}_{emN}^{(1)}(k_{1s}),\mathbf{N}_{om0}^{(1)}(k_{1s}) \rangle & \dots & \langle \mathbf{M}_{emN}^{(1)}(k_{1s}),\mathbf{M}_{emN}^{(1)}(k_{1s})
\end{bmatrix} \mathcal{C}_{i}^{m}=\begin{bmatrix}
\langle \mathbf{L}_{om0}^{(1)}(k_{1l}),\mathbf{u}_i \rangle \\
\langle \mathbf{N}_{om0}^{(1)}(k_{1l}),\mathbf{u}_i \rangle \\
\dots \\
\langle \mathbf{M}_{emN}^{(1)}(k_{1l}),\mathbf{u}_i \rangle
\end{bmatrix}.
\end{align}
The matrix on the LHS is the projection of the elements of vector $\mathcal{P}_{i}^{m}$ onto itself, and the vector on the RHS is the projection of the incident field onto $\mathcal{P}_{i}^{m}$. Note that in $\mathcal{P}_{i}^{m}$ all the VSHs are calculated with wavenumbers $k_{1s}$ and $k_{1l}$. 

The 1st boundary condition (continuity of the displacement) can be thus expressed as 
\begin{equation}
(\mathcal{P}_{\textrm{un}}^m)^T \mathcal{C}_{\textrm{un}}^{m} = ( \mathcal{P}_{i}^m)^T \mathcal{C}_{i}^{m}.
\end{equation}
To arrive at a convenient expression for coefficients in $\mathcal{C}_{\textrm{un}}^{m}$, we multiply both sides of the equation from the left by the $6(N+1)$ elements of the vector $(\mathcal{P}_{i}^{m})^*$, and integrate them over the surface of a sphere (as in our definition of the scalar product).

\begin{align}\label{bc1.arithmetic}
&\stackMath\def\stackalignment{r}%
\stackon%
{6(N+1)\mathrm{~rows}\left\{\tmp\right.}%
{\overbrace{\phantom{\smash{\tmp\mkern -36mu}}}^{12(N+1) \mathrm{\textstyle ~columns}}\mkern 20mu} \mathcal{C}_{\textrm{un}}^{m} \nonumber\\
&=\stackMath\def\stackalignment{r}%
\stackon%
{\tmpp}%
{\overbrace{\phantom{\smash{\tmpp\mkern -36mu}}}^{6(N+1) \mathrm{\textstyle ~columns}}\mkern 20mu} \mathcal{C}_{i}^{m}%
\end{align}
This equation gives us $6(N+1)$ equations for the elements of  $\mathcal{C}_{\textrm{un}}^{m}$ (which has $12(N+1)$ elements).

\subsubsection{Representing the second BCs for the fields decomposed into VSHs}

To get the remaining $6(N+1)$ equations, we turn to the 2nd boundary condition. For each of the vector spherical harmonics for the set of VSHs, we can calculate the associated radial component of the stress, using operator $\mathcal{S}$ defined through
\begin{equation}
\mathcal{S}_r[\mathbf{u}] = (T_{rr}, T_{r \theta}, T_{r \phi}),
\end{equation}
with the stress tensor elements given in Eqs.~(2) and (3). 
The material properties ($\rho_{1}$, $v_{s,1}$, $\sigma_1$, $\mu_1$ or $\rho_{2}$, $v_{s,2}$, $\sigma_2$, $\mu_2$) have to be chosen appropriately.

The algebraic expression of the 2nd boundary condition is constructed in a similar manner as the 1st BC (Eq.~(\ref{bc1.arithmetic})), but with the radial stress operator applied to all the right terms in the expressions for the scalar product:

\begin{align}\label{bc2.arithmetic}
&\stackMath\def\stackalignment{r}%
\stackon%
{\tmpS}%
{\phantom{\smash{\tmpS\mkern -36mu}}\mkern 20mu} \mathcal{C}_{\textrm{un}}^{m} \nonumber\\
&=\stackMath\def\stackalignment{r}%
\stackon%
{\tmppS}%
{\phantom{\smash{\tmppS\mkern -36mu}}\mkern 20mu} \mathcal{C}_{i}^{m}.%
\end{align}

\subsubsection{Quasinormal modes}\label{Appendox.QNMs}
Collecting the matrices from the LHSs of Eqs.~(\ref{bc1.arithmetic}) and (\ref{bc2.arithmetic}), we can write down explicitly the entire $12(N+1)\times 12(N+1)$ matrix describing the two boundary conditions:
\begin{align}\label{bc.both}
\mathcal{B}_{\textrm{un}}^m=\stackMath\def\stackalignment{r}%
\stackon%
{\tmpboth}%
{\phantom{\smash{\tmpboth\mkern -36mu}}\mkern 20mu}.
\end{align}
The quasinormal modes (QNMs) of the system are thus identified as vectors of expansion coefficients $\mathcal{N}^m$ describing the field inside and outside the sphere, such that 
\begin{equation}\label{normal.modes}
\mathcal{B}_{\textrm{un}}^m \mathcal{N}^m = 0,
\end{equation}
i.e. solving the boundary conditions in the absence of external excitation. Since the BCs do not mix different azimuthal orders $m$, normal modes can be found separately for each $m$.

We have implemented a numerical solver of Eq.~(\ref{normal.modes}) in which the complex frequency space $\tilde{\Omega}$ is explored to find the zeros of the determinant of $\mathcal{B}_{\textrm{un}}^m(\tilde{\Omega})$. This step becomes significantly more robust when the values in the part of $\mathcal{B}_{\textrm{un}}^m$ corresponding to the second BS are rescaled to the same order of magnitude as the values in the first $6(N+1)$ rows of $\mathcal{B}_{\textrm{un}}^m$, and when empty rows and columns are removed, e.g. elements corresponding to the products of $n=0$ VSHs for $\mathcal{B}_{\textrm{un}}^{m=1}$. As we approach such complex frequencies yielding $\det \mathcal{B}_{\textrm{un}}^m=0$ with arbitrary precision, the QNMs are then found as the eigenvectors of $\mathcal{B}_{\textrm{un}}^m$ corresponding to the smallest absolute eigenvalue.

\subsection{Scattering efficiency}\label{App.scattering}
In Fig. 2 we present the scattering efficiency spectra of two chosen elastic spheres. The scattering efficiency $C_{\text{sca}}$ is, as in electromagnetic scattering theory,\cite{bohren2008absorption} defined as the ratio between scattering cross section $\sigma_{\text{sca}}$ (measured in $[m^2]$), and the geometric cross section $\pi a^2$. 

The cross-section $\sigma_{\text{sca},\alpha}$ for scattering of a longitudinal ($\alpha=l$) or transverse ($\alpha=s$) wave is calculated by integrating the time-averaged flux of the Poynting vector of the scattered wave, defined in Eq.~(\ref{P.vector}), through a large sphere $\mathcal{C}(0,R)$:
\begin{equation}\label{scattering.eff.def}
\sigma_{\text{sca},\alpha} = \frac{\int_{\mathcal{C}(0,R)} \mathbf{P}\cdot \mathbf{n} ~d\sigma}{P_{\text{inc},\alpha}} = \frac{\int_{\mathcal{C}(0,R)} P_r ~d\sigma}{P_{\text{inc},\alpha}},
\end{equation}
where $P_{\text{inc},\alpha}$ is the only non-vanishing component of the Poynting vector of the incident longitudinal or transverse plane wave with amplitude $u_0$ and frequency $\Omega$, propagating in a homogeneous isotropic medium with longitudinal or transverse velocities $v_{1l}$ and $v_{1s}$, respectively:
\begin{equation}
P_{\text{inc},l} = \frac{\Omega^2}{2}\rho_1 v_{1l}|u_0|^2, \quad P_{\text{inc},s} = \frac{\Omega^2}{2}\rho_1 v_{1s}|u_0|^2.
\end{equation}

To calculate the radial component of the Poynting vector $P_r$, we make use of the expansion of the matrix introduced earlier, containing the scalar products of the VSHs with  radial components of the stress tensor associated with each of the VSHs (see the LHS of Eq.~(\ref{bc2.arithmetic})). It should be noted that this matrix has to be modified to use the spherical Hankel functions $h_n^{(1)}$ rather than spherical Bessel functions $j_n$, and that the scalar products are calculated for the radial coordinate $r=R$ instead of $r=a$. By multiplying this matrix on the right by a vector of expansion coefficients describing the scattered field, and on left by the hermitian conjugate of this vector, we obtain the integrated radial component of the vector
\begin{equation}
\int_{\mathcal{C}(0,R)} \left(u_r^* T_{rr}+u_{\theta}^* T_{\theta r}+u_{\phi}^* T_{\phi r}\right) d\sigma.
\end{equation}
This magnitude is related to $P_r$ via Eq.~(\ref{radial.Poynting}), and so the scattering cross section is given by
\begin{equation}\label{scattering.eff.def2}
\sigma_{\text{sca},\alpha} = \frac{\frac{\Omega}{2}\Im \left[\int_{\mathcal{C}(0,R)} \left(u_r^* T_{rr}+u_{\theta}^* T_{\theta r}+u_{\phi}^* T_{\phi r}\right) d\sigma\right]}{\frac{\Omega^2}{2} \rho_1 v_{1\alpha} |u_0|^2} = \frac{\Im\left[\int_{\mathcal{C}(0,R)}\left(u_r^* T_{rr}+u_{\theta}^* T_{\theta r}+u_{\phi}^* T_{\phi r}\right) d\sigma\right]}{\Omega \rho_1 v_{1\alpha} |u_0|^2}.
\end{equation}
It should be noted that since the Poynting vector decays with the square of the radial coordinate, the above results should be independent of the radius of integration sphere.

\section{Oscillating linear force in a homogeneous, isotropic medium}\label{App.radiation}

\subsection{Elastic radiation}

Radiation pattern from an oscillating linear force can be calculated from the elastic analogue of the time-averaged Poynting vector
\begin{equation}\label{P.vector}
\mathbf{P}(\mathbf{r}) = -\frac{1}{2}\Re\left[\mathbf{v}^*(\mathbf{r})\cdot \mathbf{T}(\mathbf{r})\right],
\end{equation}
where $\mathbf{v}=\partial_t \mathbf{u} = -i \Omega \mathbf{u}$ is the local velocity of displacement field and $\mathbf{T}$ is the stress tensor.
Without losing generality, we can position the oscillating force at the center of the coordinate system, and limit our attention to the radial component of the Poynting vector, arriving at:
\begin{equation}\label{radial.Poynting}
P_r = -\frac{1}{2}\Re(v_r^* T_{rr}+v_{\theta}^* T_{\theta r}+v_{\phi}^* T_{\phi r}) = \frac{\Omega}{2}\Im(u_r^* T_{rr}+u_{\theta}^* T_{\theta r}+u_{\phi}^* T_{\phi r}).
\end{equation}
In spherical coordinates, and isotropic homogeneous medium, the stress components are given by Eqs.~(2) and (3) in the main text. For the force oriented along the $\hat{\mathbf{z}}$ axis, the displacement has no $\phi$ component ($u_{\phi}=0$), and is $\phi$-independent ($\partial_{\phi}\mathbf{u}=0$).  Consequently, the expressions for the elements of the stress tensor simplify to
\begin{equation}\label{stress.def.App.2}
T_{rr} = 2 \rho v_s^2\left(\frac{\sigma}{1-2\sigma}\left\{\frac{1}{r^2}\frac{\partial}{\partial r}(r^2 u_r) + \frac{1}{r \sin \theta} \frac{\partial}{\partial \theta} [u_\theta \sin \theta]\right\} + \frac{\partial u_r}{\partial r}\right),
\end{equation}
\begin{equation}\label{stress.def.App.2b}
T_{r\theta} = T_{\theta r} = \mu \left[\frac{\partial u_\theta}{\partial r} - \frac{u_\theta}{r}+ \frac{1}{r}\frac{\partial u_r}{\partial \theta}\right],\quad T_{r\phi}=T_{\phi r}=0,
\end{equation}
where $\sigma$ is the Poisson ratio, and $v_s$ is the velocity of a shear wave. The radial component of the Poynting vector in the far-field ($\textrm{FF}$) $P_{r,\textrm{FF}}$ is determined from the corresponding ($\textrm{FF}$) part of the displacement field generated by the emitter. For the emitter $\mathbf{f}$ positioned at the center of the coordinate system $\mathbf{r}_0=(0,0,0)$, and aligned with axis $\hat{\mathbf{z}}$, the displacement field can be found from $\mathbf{u}(\mathbf{r}) = \mathbf{G}(\mathbf{r},\mathbf{r}_0) \mathbf{f}(\mathbf{r}_0)$ and the definition of the Green function given in Eq.~(5)
\begin{align}\label{uFF}
\mathbf{u}_{\textrm{FF}}(\mathbf{r}) = &\left[\frac{e^{i k_l r}}{4 r \pi (\lambda + 2\mu)}\hat{\mathbf{r}}\hat{\mathbf{r}}^T+\frac{e^{i k_s r}}{4 r \pi \mu}\left(\mathbbm{1}-\hat{\mathbf{r}}\hat{\mathbf{r}}^T\right)\right] \mathbf{f} =\frac{e^{i k_l r}}{4 r \pi (\lambda + 2\mu)}\hat{\mathbf{r}}\cos \theta  f-\frac{e^{i k_s r}}{4 r \pi \mu}\hat{\mathbf{\theta}}\sin \theta f,
\end{align}
where we used $\hat{\mathbf{z}} = \hat{\mathbf{r}}\cos \theta  - \hat{\mathbf{\theta}}\sin \theta$, and introduced the longitudinal ($k_l=\Omega/v_l$) and shear ($k_s=\Omega/v_s$) wavenumbers. The expressions for the spherical Hankel functions of the first kind $h_0^{(1)}$ and $h_2^{(1)}$ are
\begin{equation}
h_0^{(1)}(x) = -i\frac{e^{ix}}{x},\quad h_2^{(1)}(x) = e^{ix}\left(\frac{i}{x}-\frac{3}{x^2} - \frac{3i}{x^3}\right).
\end{equation}
Using Eqs.~(\ref{stress.def.App.2}-\ref{uFF}), we can write the radial component of the Poynting vector in the far field as
\begin{align}
P_{r,\textrm{FF}}(\mathbf{r})\frac{2}{\Omega |f|^2}
= &
\Im\left(\frac{\cos^2 \theta \rho v_s^2}{8 r^3 \pi^2 (\lambda + 2\mu)}\left\{\frac{1}{(\lambda + 2\mu)} \left[\frac{\sigma}{1-2\sigma} (1+ i k_l r) + (i k_l r -1)\right] \right\} + \frac{\sin^2 \theta}{16 r^3 \pi^2 \mu}(i k_s r -2) \right.\\ \nonumber
&\left. + \frac{\cos^2 \theta \rho v_s^2}{8 r^3 \pi^2 (\lambda + 2\mu)}\left[-\frac{\sigma}{1-2\sigma}
\frac{2 e^{i (k_s-k_l) r}}{\mu} \right] +\frac{\sin^2 \theta}{16 r^3 \pi^2} \frac{e^{i (k_l-k_s) r}}{(\lambda + 2\mu)}\right).
\end{align}
The first two contributions describe the far-field Poynting vector in the far field associated with transverse and longitudinal waves, respectively, decaying as $r^{-2}$. The last two terms describe interference between the two waves, and decay as $r^{-3}$. Thus, we can eliminate them, finding the radiation pattern
\begin{align}
P_{r,\textrm{FF}}(\mathbf{r})\frac{2}{\Omega |f|^2} \rightarrow
&
\frac{1}{16 r^2 \pi^2}\left(\frac{k_l}{\lambda + 2\mu} \cos^2 \theta + \frac{k_s}{\mu}  \sin^2 \theta\right),
\end{align}
where we used the property of Poisson ratio $\sigma=\lambda/(2(\lambda + \mu))$,\cite{auld1973acoustic} and the definition of the shear velocity $v_s = \sqrt{{E}/({2\rho(1+\sigma)})}$.

Therefore, in the far-field regime, the interference between the longitudinal and transverse waves vanishes. The radiation pattern of the transverse waves is given by the $\sin^2\theta$ function, recognized as that of an electromagnetic dipolar source. For the longitudinal waves, the radiation pattern follows the $\cos^2\theta$ distribution. These features are shown schematically in Fig. 1(b).

Furthermore, we note that by integrating this energy flux over the surface of a large sphere, we can find the radiation emission power of the localized elastic emitter. In the absence of losses, this magnitude is
\begin{align}\label{power.P0}
P_0 &= \int_{0}^\pi d\theta \int_0^{2\pi} d\phi~r^2 \sin \theta \frac{\Omega |f|^2}{32 r^2 \pi^2}\left(\frac{k_l}{\lambda + 2\mu} \cos^2 \theta + \frac{k_s}{\mu}  \sin^2 \theta\right)= \frac{\Omega |f|^2 }{12\pi}\left[\frac{k_l}{2 (\lambda + 2\mu)} + \frac{k_s}{\mu}\right].
\end{align}

\subsection{Local partial density of elastic states in a homogeneous medium}\label{App.G00}

Power radiated by an elastic emitter in an isotropic, homogeneous environment, can be alternatively obtained by considering the uniform medium Green function $\mathbf{G}(\mathbf{r},\mathbf{r}_0)$ in the limit of $\mathbf{r}\rightarrow\mathbf{r}_0$, as described by Eq.~(4) of the main text. We can consider approaching the $\mathbf{r_0}$ along the axis of $\mathbf{f}$. The power radiated through the shear components is:
\begin{equation}\label{P0S}
P_{0,s} = \frac{\Omega}{2} \Im\left\{-\frac{i k_s}{12 \pi \mu} |f|^2 \left[ -2 h_0^{(1)}(d)- 2 h_2^{(1)}(d) \right]\right\} = \frac{\Omega}{2} \Im\left\{\frac{i k_s}{2 \pi \mu} |f|^2 e^{ik_s d} \frac{1}{(k_s d)^2} \left[ -1 - \frac{i}{k_s d} \right]\right\},
\end{equation}
where $d = |\mathbf{r}-\mathbf{r}_0|$. Following the case for EM, we expand the exponents as
\begin{equation}
\exp(ik d) \approx 1 + ikd - \frac{1}{2}(kd)^2 - \frac{i}{6}(kd)^3.
\end{equation}
The product of the exponent and the expression in the square brackets is then
\begin{align}
&\left[1 + ik_sd - \frac{1}{2}(k_sd)^2 - \frac{i}{6}(k_sd)^3\right] \left[ -1 - \frac{i}{k_s d} \right]= \left[-\frac{i}{k_s d} -\frac{i}{2}k_sd+ \frac{1}{3}(k_sd)^2 + \frac{i}{6}(k_sd)^3\right],
\end{align}
and thus we get
\begin{equation}
P_{0,s} = \frac{\Omega |f|^2 }{12 \pi \mu}k_s.
\end{equation}

We can conduct a similar analysis for the longitudinal waves component of the Green function:
\begin{equation}
P_{0,l} = \frac{\Omega}{2} \Im\left\{\frac{i k_l}{12 \pi (\lambda + 2\mu)}|f|^2 \left[ h_0^{(1)}(k_l d)- 2 h_2^{(1)}(k_l d) \right]\right\} = \frac{\Omega}{2} \Im\left\{\frac{i k_l}{2 \pi (\lambda + 2\mu)}\frac{|f|^2}{(k_l d)^2}  e^{ik_l d} \left[ -\frac{i}{2}k_l d + 1+ \frac{i}{k_l d}\right]\right\}.
\end{equation}
Applying the same expansion of the exponent as before, we find two imaginary terms, rather than one as in Eq. (\ref{P0S}), proportional to $(k_l d)^2$ and $(k_l d)^4$. Since the second quartic term gives rise to a contribution vanishing with $d\rightarrow 0$, we can find the power radiated through the longitudinal waves is given by
\begin{equation}
P_{0,l} = \frac{\Omega |f|^2 }{24 \pi (\lambda + 2\mu)}k_l.
\end{equation}
The sum of the powers $P_{0,l}$ and $P_{0,s}$ gives us thus identical expression to that derived by integrating the flow of the Poynting vector (Eq.~(\ref{power.P0})).

{
	\section{Derivation of elastic Purcell factor}\label{Appendix.Purcell}
	In this Appendix we derive the definitions of the elastic Purcell factor and the elastic effective mode volume given in the main text. The analogous derivations conducted for the optical case in Ref. \cite{Koenderink:10,esteban2014strong,PhysRevB.91.195422} often deal with the much more difficult case of Purcell factor in dispersive, absorbing media. While we chose an elastic setup which does not suffer from these problems, we still need to carefully consider the radiative losses related to the elastic radiation emitted from the QNMs of the cavity.
	
	
	The formal quantization of the modes of an elastic cavity was presented in Ref. \citenum{sipe2016hamiltonian}. The displacement operator associated with the $\alpha$ mode is in this picture given by:
	\begin{equation}
	\hat{\mathbf{u}}_\alpha(\mathbf{r}) = \sqrt{\frac{\hbar}{2 \Omega_\alpha}} \left[\hat{b}_\alpha \frac{\mathbf{U}_\alpha(\mathbf{r})}{\sqrt{\rho(\mathbf{r})}} + \hat{b}_\alpha^\dag\frac{ \mathbf{U}_\alpha^*(\mathbf{r})}{\sqrt{\rho(\mathbf{r})}}\right].
	\end{equation}
	Here $\hat{b}_\alpha$ and $\hat{b}_\alpha^\dag$ are the annihilation and creation operators for a cavity mode $\alpha$ with frequency $\Omega_\alpha$ and radiative decay rate $\gamma_\alpha$. $\mathbf{U}_\alpha(\mathbf{r})$ is the normalized mode profile which satisfies the orthonormality property
	\begin{equation}\label{ortho}
	\int \mathrm{d}\mathbf{r}~\mathbf{U}_\alpha^*(\mathbf{r}) \cdot \mathbf{U}_{\alpha'}(\mathbf{r}) = \delta_{\alpha \alpha'}.
	\end{equation}
	It should thus be noted that since the orthogonality definition does not include the density field, displacement fields $\mathbf{u}_\alpha$ associated with cavity modes do not, in general, form an orthonormal set. Nevertheless, this normalisation will give us later a direct path to formulating the effective volume of an elastic mode.
	
	
	The elastic emitter will be modeled as a two-level system with transition operators $\hat{\mathbf{f}} = \tilde{\mathbf{f}} (\ket{g}\bra{e} + \ket{e}\bra{g})$, with the ground and excited state denoted as $\ket{g}$ and $\ket{e}$, respectively, separated in energy by $\hbar \Omega_m$. We should note that the transition moment $\tilde{\mathbf{f}}$ is related to the classical force ${\mathbf{f}}$ via $\mathbf{f}=2\tilde{\mathbf{f}}$. Building on this quantum-classical analogy, we can derive the rate of the spontaneous transitions from the excited to the ground state for the emitter positioned in a homogeneous medium $\gamma_0 \hbar \Omega_m =  P_0$.
	
	To calculate the rate of the spontaneous emission from an emitter positioned at $\mathbf{r}_m$ into a cavity mode $\alpha$, we consider the interaction Hamiltonian
	\begin{equation}
	\hat{H}_I = \hat{\mathbf{f}}\cdot \hat{\mathbf{u}}_\alpha(\mathbf{r}_m) = \sqrt{\frac{\hbar}{2 \Omega_\alpha \rho(\mathbf{r}_m)}}  (\ket{g}\bra{e} + \ket{e}\bra{g}) \tilde{\mathbf{f}}\cdot\left[\hat{b}_\alpha \mathbf{U}_\alpha(\mathbf{r}_m)+\hat{b}_\alpha^\dag \mathbf{U}^*_\alpha(\mathbf{r}_m)\right].
	\end{equation}
	In the rotating wave approximation, this simplifies to
	\begin{equation}
	\hat{H}_{I,\textrm{rwa}} = \sqrt{\frac{\hbar}{2 \Omega_\alpha \rho(\mathbf{r}_m)}}  \tilde{\mathbf{f}}\cdot \mathbf{U}_\alpha(\mathbf{r}_m) \hat{b}_\alpha \ket{e}\bra{g} + h.c. = \hbar g_\alpha \hat{b}_\alpha \ket{e}\bra{g} + h.c.,
	\end{equation}
	where we introduced the coupling constant $g_\alpha$.
	The rate of spontaneous emission $\gamma_\alpha$ into the $\alpha$ cavity mode is thus given as\cite{esteban2014strong,auffeves2010controlling}
	\begin{equation}\label{gamma.alpha}
	\gamma_\alpha = \frac{4 g_\alpha^2}{\kappa_\alpha} = \frac{2}{\hbar \kappa_\alpha \Omega_\alpha \rho(\mathbf{r}_m)} |\tilde{\mathbf{f}}\cdot \mathbf{U}_\alpha(\mathbf{r}_m)|^2,
	\end{equation}
	where $\kappa_\alpha$ is the decay rate of the $\alpha$ mode. If the emitter has the same polarization as the displacement field (so that $\tilde{\mathbf{f}}\cdot \mathbf{U}_\alpha(\mathbf{r}_m) = \tilde{f} U_\alpha(\mathbf{r}_m)$), we get
	\begin{equation}\label{blah1}
	\gamma_\alpha = \frac{2}{\hbar\kappa_\alpha \Omega_\alpha \rho(\mathbf{r}_m)} |\tilde{f}|^2 |U_\alpha(\mathbf{r}_m)|^2.
	\end{equation}
	Before we continue, let us consider two aspects of the Purcell factor:
	\begin{itemize}
		\item the Purcell factor is defined as the enhancement of the spontaneous decay rate, and so it is given by the ratio of $\gamma_\alpha$ and the rate of spontaneous emission for the emitter decoupled from the cavity modes $\gamma_0$; as we have discussed earlier, this rate depends on the material properties of the homogeneous, infinite medium in which the emitter is positioned,
		\item simultaneously, the Purcell factor was originally defined solely in terms of the properties of a single cavity mode, and did not include any details quantifying the positioning or the orientation of the emitter in the cavity;\cite{Purcell1946} rather, it was implicitly assumed that the emitter would be positioned and oriented \textit{optimally}, so that the coupling would be maximized, and the Purcell factor would serve as an upper limit for the decay rate enhancement in this case. 
	\end{itemize}
	These two conditions are straightforward and consistent in typical microcavity electromagnetic or plasmonic setups. However, they become problematic if we consider the elastic resonators discussed in this Letter, where the \textit{maximum coupling} is found for the emitter positioned inside the antenna. Therefore, the normalizing spontaneous emission rate $\gamma_0$ should be calculated assuming that the emitter is immersed in the material making up the antenna, rather than the surrounding medium. This picture presents a number of problems, e.g.: (i) it does not correspond to any useful physical picture of the antenna as a device that can be used to modify the radiation from nearby emitters, (ii) one could conceive an setup where the maximized $\gamma_\alpha$ obtained for the emitter positioned inside the antenna does not yield the maximum Purcell factor due to the large contrast of $\gamma_0$'s in the medium and the antenna. 
	
	An alternative definition of the Purcell factor, which we use here, relies on normalizing $\gamma_\alpha$ by the spontaneous emission rate $\gamma_0$ calculated in the medium:
	\begin{align}\label{Purcell.2}
	F_\alpha &= \frac{\gamma_\alpha}{\gamma_0} = \gamma_\alpha \frac{\hbar \Omega_m}{P_0} \\ \nonumber
	&= \frac{2}{\hbar\kappa_\alpha \Omega_\alpha \rho(\mathbf{R}_M)} |\tilde{f}|^2 |U_\alpha(\mathbf{R}_M)|^2 \frac{12 \pi \hbar}{|f|^2}\left[\frac{k_{s,1}}{\mu_1} + \frac{k_{l,1}}{2(\lambda_1 + 2\mu_1)}\right]^{-1} 
	= \frac{6\pi Q_\alpha}{\Omega_\alpha^2 \Omega_m \rho(\mathbf{R}_M) \rho_1^{1/2}\phi_1} |U_\alpha(\mathbf{R}_M)|^2,
	\end{align}
	where we used the definition of $P_0$ introduced in Eq.~(\ref{power.P0}) and the definitions of the shear and longitudinal velocities in the environment. We also introduced the quality factor of the resonator $Q_{\alpha} = \Omega_\alpha/\kappa_\alpha$ and the coordinate $\mathbf{R}_M$ denoting the position of the emitter providing the largest coupling to the cavity mode, i.e.
	\begin{eqnarray}\label{Veff}
	\max_{\mathbf{r}}\left[|U_\alpha(\mathbf{r})|^2\right] = |U_\alpha(\mathbf{R}_M)|^2 = V_{\alpha,\text{eff}}^{-1}.
	\end{eqnarray}
	The last identity should be treated as a definition of the effective volume of the elastic mode $\alpha$. While this formulation is very simple, estimation of $V_{\alpha,\text{eff}}$ requires the knowledge of the mode profiles $U_\alpha$. To this end, we recall the definition of displacement operator $\hat{\mathbf{u}}_\alpha(\mathbf{r})$, and consider its positive-frequency, classical version:
	\begin{eqnarray}
	\mathbf{u}_\alpha(\mathbf{r}) = \sqrt{\frac{\hbar}{2 \Omega_\alpha \rho(\mathbf{r})}}\mathbf{U}_\alpha(\mathbf{r}).
	\end{eqnarray}
	We describe briefly the protocol for finding the displacement fields $\mathbf{u}_\alpha$ related to the QNMs in Appendix \ref{Appendox.QNMs}. Consequently, to ensure the proper normalization (Eq.~(\ref{ortho})) of the mode profiles $\mathbf{U}_\alpha$, we define them as:
	\begin{eqnarray}\label{U}
	\mathbf{U}_\alpha(\mathbf{r}) = \frac{\sqrt{\rho(\mathbf{r})} \mathbf{u}_\alpha(\mathbf{r})}{\sqrt{\int d\mathbf{r}' \rho(\mathbf{r}') |{\mathbf{u}_\alpha}(\mathbf{r}')|^2}}.
	\end{eqnarray}
	The effective mode volume can thus be expressed as
	\begin{eqnarray}
	V_{\alpha,\text{eff}} = \frac{\int d\mathbf{r}' \rho(\mathbf{r}') |{\mathbf{u}_\alpha}(\mathbf{r}')|^2}{\left[\rho(\mathbf{R}_M) |{\mathbf{u}_\alpha}(\mathbf{R}_M)|^2\right]}.
	\end{eqnarray}
	Finally, assuming that $\mathbf{R}_M$ is found inside the antenna ($\rho(\mathbf{R}_M) = \rho_2$), for the emitter tuned to the resonant frequency of the cavity mode ($\Omega_m = \Omega_{\alpha}$), we use the identity in Eq.~(\ref{Veff}) and get
	\begin{align}\label{F3}
	F_\alpha &= \frac{6\pi Q_\alpha}{\Omega_\alpha^3 \rho_2 \rho_1^{1/2}\phi_1 V_{\alpha,\text{eff}}}.
	\end{align}
	
	\subsection{Modified Purcell factor}
	The definition of the Purcell factor can be extended, as discussed in the letter, to account for the dependence on the orientation, position and detuning of the emitter with respect to the cavity resonance $\Omega_\alpha$. To this end, we recall Eqs.~(\ref{Purcell.2}) without assuming optimal orientation of the emitter, and additionally, replace the optimal position of the emitter $\mathbf{R}_M$ with its actual position $\mathbf{r}_m$. The modified Purcell factor is then
	\begin{equation}
	\tilde{F}_{\alpha}(\mathbf{r}_m) = \frac{6\pi Q_\alpha}{\Omega_\alpha^2 \Omega_m \rho(\mathbf{r}_m) \rho_1^{1/2}\phi_1} \frac{|\tilde{\mathbf{f}} \cdot \mathbf{U}_\alpha(\mathbf{r}_m)|^2}{|\tilde{\mathbf{f}}|^2}.
	\end{equation}
	The last fraction can be expressed using Eq.~(\ref{U}):
	\begin{equation}
	|\tilde{\mathbf{f}} \cdot \mathbf{U}_\alpha(\mathbf{r}_m)|^2 = \frac{\rho(\mathbf{r}_m) |\tilde{\mathbf{f}} \cdot \mathbf{u}_\alpha(\mathbf{r}_m)|^2}{\int d\mathbf{r}' \rho(\mathbf{r}') |{\mathbf{u}_\alpha}(\mathbf{r}')|^2} = \frac{1}{V_{\alpha,\text{eff}}}\frac{\rho(\mathbf{r}_m) |\tilde{\mathbf{f}} \cdot \mathbf{u}_\alpha(\mathbf{r}_m)|^2}{\rho(\mathbf{R}_M) |\mathbf{u}_\alpha(\mathbf{R}_M)|^2}.
	\end{equation}
	The detuning dependence is introduced through lorentzian profile of the cavity mode of with $\kappa_\alpha$. The modified Purcell factor can be finally related to the definition of the Purcell factor given in Eq.~(\ref{F3})
	\begin{align}
	\tilde{F}_{\alpha}(\mathbf{r}_m) 
	= F_{\alpha} \frac{\Omega_\alpha}{\Omega_m} \frac{|\tilde{\mathbf{f}} \cdot \mathbf{u}_\alpha(\mathbf{r}_m)|^2}{ |\tilde{\mathbf{f}}|^2|\mathbf{u}_\alpha(\mathbf{R}_M)|^2 }\frac{(\kappa_\alpha/2)^2}{(\Omega_\alpha- \Omega_m)^2+(\kappa_\alpha/2)^2}.
	\end{align}
	
}

\bibliographystyle{apsrev4-1} 
\bibliography{bibliography}

\begin{thebibliography}{45}
\bibitem [{Not()}]{Notation}%
  \BibitemOpen
  \href@noop {} {}\bibinfo {note} {According to the conventional notation, the
  second subscript $l$ numbers the roots of the characteristic equation. For
  shear modes for $n\neq 0$, $l-1$ is replaced by $l$.}\BibitemShut {Stop}%
\end{thebibliography}

\end{document}